\documentclass{article}
\usepackage{PRIMEarxiv}
\usepackage{tikz}

\usepackage{amssymb}
\usepackage{array}
\usepackage{hyperref}
\usepackage{orcidlink}
\usepackage{balance}
\usepackage{booktabs}
\usepackage{siunitx}
\usepackage{siunitx}
\usepackage{multirow}
\usepackage{graphicx}
\usepackage{xspace}
\usepackage[table]{xcolor}
\usepackage[linesnumbered,ruled,vlined]{algorithm2e}
\usepackage{algpseudocode}
\usepackage[most]{tcolorbox}

\usepackage[utf8]{inputenc} % allow utf-8 input
\usepackage[T1]{fontenc}    % use 8-bit T1 fonts
\usepackage{hyperref}       % hyperlinks
\usepackage{url}            % simple URL typesetting
\usepackage{booktabs}       % professional-quality tables
\usepackage{amsfonts}       % blackboard math symbols
\usepackage{nicefrac}       % compact symbols for 1/2, etc.
\usepackage{microtype}      % microtypography
\usepackage{lipsum}
\usepackage{graphicx}
\graphicspath{ {./images/} }

\newcommand{\blackcircled}[1]{\tikz[baseline=(char.base)]{
            \node[shape=circle,fill=black,inner sep=0.7pt] (char) {\textcolor{white}{#1}};}}

\newcommand{\whitecircled}[1]{\tikz[baseline=(char.base)]{
            \node[shape=circle,draw=black,fill=white,inner sep=0pt, minimum size=4pt] (char) {\textcolor{black}{#1}};}}
\newcommand{\relrepair}{\textsc{RelRepair}\xspace}

\AtBeginDocument{%
  }

\title{RelRepair: Enhancing Automated Program Repair by Retrieving Relevant Code}

\author{
 Shunyu Liu \\
  The University of Queensland \\
  Brisbane, Australia \\
  \texttt{shunyu.liu@uq.edu.au}
  \And
 Guangdong Bai \\
  The University of Queensland \\
  Brisbane, Australia \\
  \texttt{g.bai@uq.edu.au}
  \And
 Mark Utting \\
  The University of Queensland \\
  Brisbane, Australia \\
  \texttt{m.utting@uq.edu.au}
  \And
 Guowei Yang \\
  The University of Queensland \\
  Brisbane, Australia \\
  \texttt{guowei.yang@uq.edu.au}
}

\date{September 20, 2025}

\begin{document}
\maketitle
\begin{abstract}

Automated Program Repair (APR) has emerged as a promising paradigm for reducing debugging time and improving the overall efficiency of software development. Recent advances in Large Language Models (LLMs) have demonstrated their potential for automated bug fixing and other software engineering tasks. Nevertheless, the general-purpose nature of LLM pre-training means these models often lack the capacity to perform \emph{project-specific repairs}, which require understanding of domain-specific identifiers, code structures, and contextual relationships within a particular codebase. As a result, LLMs may struggle to generate correct patches when the repair depends on project-specific information.
% \bai{I suggest we not attribute this to accessibility, but to the nature of general purpose oriented training, which makes them incapable of domain-specific task.}

To address this limitation, we introduce \relrepair, a novel approach that retrieves relevant 
project-specific code to enhance automated program repair.
%through a specialized RAG approach tailored to program repair tasks.
%Relevance-Guided Program Repair framework built on Retrieval-Augmented Generation (RAG). 
\relrepair first identifies relevant function signatures by analyzing function names and code comments within the project. It then conducts deeper code analysis to retrieve code snippets relevant to the repair context. The retrieved relevant information is then incorporated into the LLM's input prompt, guiding the model to generate more accurate and informed patches. We evaluate \relrepair on two widely studied datasets, Defects4J V1.2 and ManySStuBs4J, and compare its performance against several state-of-the-art LLM-based APR approaches. \relrepair successfully repairs 101 bugs in Defects4J V1.2. Furthermore, \relrepair achieves a 17.1\% improvement in the ManySStuBs4J dataset, increasing the overall fix rate to 48.3\%. These results highlight the importance of providing relevant project-specific information to LLMs, shedding light on effective strategies for leveraging LLMs in APR tasks. 

\end{abstract}


\section{Introduction}
Automated Program Repair (APR) is an emerging field focused on automating the bug-fixing process in software development, significantly reducing debugging time and improving software reliability. Traditional APR methods generally operate based on a Generation-Validation (G\&V) \cite{long2016analysis} paradigm. These techniques are categorized into four main categories: heuristic-based, constraint-based, template-based, and learning-based approaches. Heuristic-based APR \cite{le2011genprog, le2016history, wen2018context} employs genetic algorithms and search heuristics to iteratively generate and evaluate patches. Constraint-based APR \cite{demarco2014automatic, long2015staged, mechtaev2016angelix, le2017s3} treats repair as a constraint satisfaction problem, solving logical formulations via program synthesis or symbolic execution. Template-based APR \cite{martinez2016astor, hua2018sketchfix, ghanbari2019practical, liu2019tbar, liu2019avatar} uses predefined patch templates that match specific bug patterns to generate fixes efficiently and accurately. Learning-based APR \cite{chen2019sequencer, jiang2021cure, li2020dlfix, lutellier2020coconut} leverages neural models, particularly Neural Machine Translation (NMT), to learn repair transformations from historical code changes. While learning-based APR achieves strong performance by capturing generalizable patterns, its effectiveness is limited by the quality %and noisiness 
of training data mined from open-source repositories.

With the rise of pre-trained Large Language Models (LLMs) whose training corpora include large volumes of open-source code,
% \bai{Check logic. They are not trained on open-source code, but have open-source code in their training dataset.}
LLM-based APR \cite{kolak2022patch, prenner2022can, xia2023automated} has emerged as a promising alternative. It employs LLMs, such as ChatGPT \cite{schulman2022chatgpt} that have been trained on billions of words from various sources, to understand the context of buggy code and directly generate its patches. This approach significantly improves the quality of repair tasks and reduces the dependence on large labeled datasets. For example, ChatRepair \cite{xia2023keep} leverages ChatGPT's conversational abilities to iteratively refine the repair process by enriching the context with information from failing tests. Similarly, ThinkRepair \cite{yin2024thinkrepair} exploits the strong analytical and reasoning capabilities of Chain-of-Thought (CoT) to deliver an advanced program-repair solution. Collectively, these recent techniques highlight the advantage of LLM-based APR, producing patches that are more 

Despite the promising performance of LLM-based APR techniques when compared to traditional methods, they still exhibit certain limitations. \textit{1)Limitations in Project-Specific Code Repair}.  While these models are proficient in handling general coding patterns due to training on broad, publicly available code corpora, they often lack familiarity with project-specific identifiers, auxiliary functions, or utility methods defined by individual developers. Consequently, generating accurate patches requires retrieving precise, relevant information directly tied to the specific bug context rather than relying solely on learned general patterns.  \textit{2) Computer Resources Challenges.} Although fine-tuning LLMs on a specific codebase could provide domain-specific adaptation, it requires considerable computational resources, making it a costly solution. To overcome these limitations, we turn to Retrieval-Augmented Generation (RAG) \cite{lewis2020retrieval}, a powerful approach that combines retrieval models with LLMs to enhance the generation process. RAG has emerged as a representative technique in the field of generative AI, 
as it effectively improves the quality of generated content by external knowledge. %RAG first employs a retriever to identify and extract relevant project-specific information. This retrieved information is then combined with the original query as additional context to guide the generation process and address the challenges faced by LLM-based APR techniques.

We propose \relrepair,
a novel approach that retrieves relevant code to
enhance automated program repair, through a specialized RAG approach tailored to program repair tasks.
%a retrieval-augmented APR framework specifically designed to improve LLM-based repair through a specialized RAG approach tailored to program repair tasks. 
RAG frameworks are inherently flexible and can be adapted to retrieve either general or highly task-specific information depending on the scenario. In our approach, \relrepair\ focuses on retrieving highly relevant, project-specific data essential for effective program repair. Specifically, it retrieves: (1) relevant function signatures, facilitating simpler repairs involving direct modifications or replacements of function signatures without extensive code analysis, and (2) relevant code snippets, capturing semantically aligned fix logic or illustrating function usage. The retrieved relevant code is incorporated into the LLM prompt at inference time, 
% enabling the model to generate \guowei{are you using "reasoning capability" of the model?} more effectively about bugs involving unfamiliar or project-specific constructs. 
allowing the model to generate patches more effectively for bugs that involve unfamiliar or project-specific constructs.
 
To support the retrieval of both function signatures and relevant code snippets, \relrepair follows a unified three-step process: (1) Query Rewriting, which reformulates search queries to improve retrieval precision; (2) Dataset Creation and Indexing, which builds a targeted index of functions and code snippets from the project; and (3) Retrieval, where similarity-based search identifies the most relevant entries. In the final patch generation phase, the retrieved information is seamlessly integrated into the LLM’s input to guide its repair decisions. This RAG-based framework enables \relrepair to generate more accurate and informed patches across diverse bug scenarios.

We evaluate \relrepair by comparing it against state-of-the-art LLM-based APR techniques on two widely studied Java benchmarks: Defects4J V1.2 \cite{just2014defects4j} and ManySStuBs4J \cite{karampatsis2020often}. 
% \bai{The introduction about the datasets, which is contained in Abstract, should be here too}
Our experiments demonstrate \relrepair's effectiveness in program repair. Specifically, in Defects4J V1.2, \relrepair successfully repairs 101 out of 255 bugs, while in ManySStuBs4J, our repair success rate improves from 31.2\% to 48.3\%. %\relrepair achieves better results than existing LLM-based methods across multiple benchmarks.

In summary, this paper makes the following contributions:
\begin{itemize}
    \item \textbf{Idea:} 
    We introduce the idea of  retrieving relevant code to enhance automated program repair. With the project-specific data that is relevant to the target bug, LLMs can potentially generate more accurate and informed patches across diverse bug scenarios.
    \item \textbf{Technique:} We apply
    a RAG framework that incorporates two complementary retrieval processes: relevant function signature retrieval and relevant code snippet retrieval, where the system first rewrites queries to optimize retrieval, then extracts function signatures or relevant code snippets from codebase. This retrieved context is integrated into LLM prompts, allowing the model to generate patches with deeper structural awareness and higher correctness.
    \item \textbf{Evaluation:} We evaluate \relrepair using two benchmarks for Java programs. The results show that \relrepair consistently fixes more bugs than LLM-based APR techniques across all evaluated benchmarks. 
\end{itemize}

\section{Background}
\subsection{Large Language Models}
Large Language Models (LLMs) \cite{brown2020language} represent a significant advancement in artificial intelligence. These models are versatile and can be applied to a range of tasks. LLMs can be adapted for specific tasks by fine-tuning \cite{radford2018improving} or prompting \cite{liu2023pre}. Fine-tuning adapts LLMs to specific tasks by further training on supplementary datasets, but this approach is resource- and time-intensive due to the need for data collection and extended training cycles. In contrast, prompting enables immediate model adaptation by providing task-specific instructions and contextual information, allowing effective performance without additional training and significantly reducing deployment time and cost.

ChatGPT~\cite{schulman2022chatgpt} is a transformer-based language model trained via a two-phase process: large-scale pretraining on diverse text to establish foundational linguistic knowledge, followed by fine-tuning with human-labeled data to better align its outputs with human preferences. This alignment is achieved by training a reward model to rank candidate responses according to human feedback~\cite{ziegler2019fine, christiano2017deep, ouyang2022training}. The model is then further refined through iterative reinforcement learning~\cite{schulman2017proximal}, enabling ChatGPT to generate responses that more closely reflect human reasoning and communication.

\subsection{Automated Program Repair}
Automated Program Repair (APR) aims to reduce manual debugging efforts by automatically generating candidate patches for faulty code, typically given a fault localization result. Traditional APR techniques can be broadly classified into three main categories: heuristic-based, constraint-based, and template-based methods. Heuristic-based APR techniques~\cite{le2011genprog, le2016history, wen2018context} utilize search-based strategies, such as genetic algorithms, to iteratively generate and evaluate program variants. Constraint-based techniques~\cite{demarco2014automatic, long2015staged, mechtaev2016angelix, le2017s3} reformulate the repair task as a constraint satisfaction or program synthesis problem, often relying on symbolic execution or SMT solvers to derive patches that satisfy semantic correctness properties. Template-based approaches~\cite{martinez2016astor, hua2018sketchfix, ghanbari2019practical, liu2019tbar, liu2019avatar} apply predefined or mined repair templates to replace or modify buggy code segments. Despite their respective strengths, these conventional APR techniques share several common limitations. Most notably, they are typically constrained to producing relatively simple patches, often one-line fixes, which limits their ability to generalize to more complex bugs. Moreover, they require substantial manual effort to adapt across programming languages, as fix patterns and repair logic must be reimplemented for each target language. 

To overcome these limitations, learning-based APR approaches leverage deep learning and large-scale open-source code to automatically learn bug-fixing patterns. Neural Machine Translation (NMT)-based models~\cite{chen2019sequencer, jiang2021cure, li2020dlfix, lutellier2020coconut} employ encoder-decoder architectures to transform buggy code into correct code, with some, like Coconut~\cite{lutellier2020coconut}, utilizing convolutional neural networks~\cite{gehring2017convolutional}. However, these methods still face challenges from noisy training data and difficulties in generalizing to diverse bug types.

Recently, the adoption of pre-trained LLMs for APR has shown promise, as LLMs can generate correct patches directly without an explicit translation step. Approaches such as ChatRepair~\cite{xia2023keep} leverage ChatGPT's conversational interface to iteratively incorporate context from failing tests, enabling broader bug coverage. Nonetheless, LLMs may still struggle to integrate project-specific functions or code snippets effectively. To address this, we propose \relrepair, an autonomous framework that systematically retrieves and supplies relevant function signatures and code snippets from the codebase, thereby enhancing LLMs' ability to generate accurate patches for complex bugs.

\subsection{Retrieval-Augmented Generation}
LLMs have exhibited remarkable performance across various natural language processing tasks, benefiting from extensive static pre-training on massive corpora. Nevertheless, their effectiveness is inherently constrained by the static nature of their embedded knowledge, limiting their ability to accurately respond to dynamic queries that require up-to-date or external information~\cite{kandpal2023large}. To overcome this limitation, Retrieval-Augmented Generation (RAG) has been proposed as a promising approach, integrating retrieval components with generative language models. At its core, RAG augments generation by retrieving task-relevant content that are semantically aligned with the input query. This retrieved information augments the LLM’s understanding, which in turn leads to more accurate and reliable patch generation. The RAG framework thus consists of three components including retrieval, generation, and augmentation ~\cite{li2022survey}.

However, traditional RAG approaches face challenges, such as retrieving irrelevant, noisy, or incomplete information, which can negatively impact the accuracy and reliability of generated outputs. Recent advances in RAG methodologies, such as those described by Gao et al.~\cite{gao2023retrieval}, introduce pre-retrieval strategies, including query rewriting. Query rewriting techniques aim to enhance retrieval precision by reformulating user queries into more targeted and precise representations. This process improves the relevance of retrieved data, ultimately leading to more accurate and effective generation. In this study, we leverage these advanced RAG strategies, integrating query rewriting to further enhance retrieval accuracy and overall model performance.

\section{MOTIVATING EXAMPLES}
To illustrate the critical role of relevant code in improving the effectiveness of bug repair, we present two motivating examples. Each example focuses on a category of information—relevant function signatures and relevant code snippets—and demonstrates how the availability or absence of such information directly influences the correctness and completeness of the patches generated by LLMs.

\subsection{Relevant Function Signatures Motivating Example}
\begin{figure}
    \centering
    \includegraphics[width=0.5\textwidth]{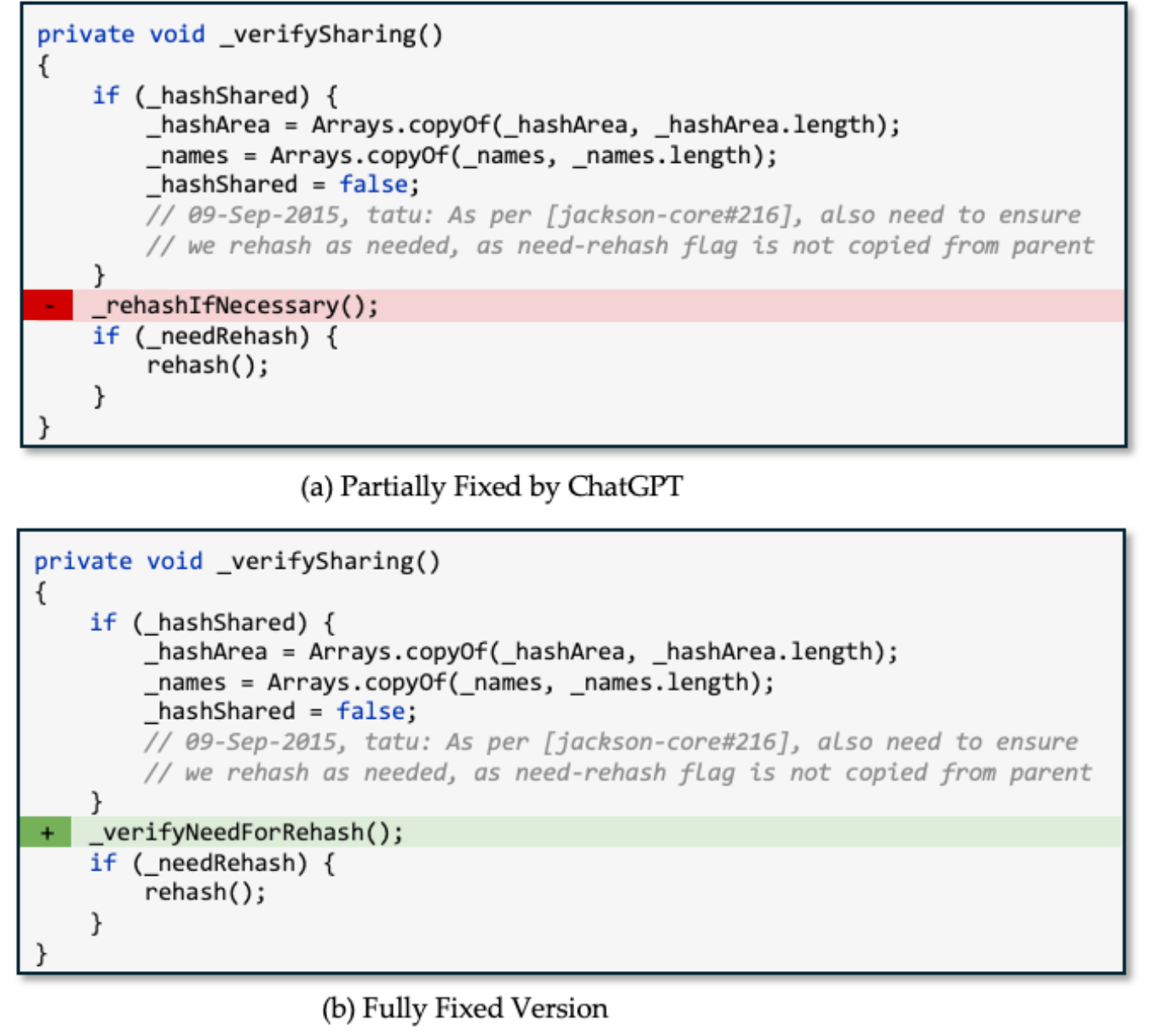}
    \caption{\centering Comparison of ChatGPT Fixed Version and Fully Fixed Version for the \href{https://github.com/FasterXML/jackson-core/issues/216}{JacksonCore-11}.}
    \label{fig:1}
\end{figure}

Figure~\ref{fig:1} presents the \href{https://github.com/FasterXML/jackson-core/issues/216}{JacksonCore-11} bug and its resolution. The original \texttt{\_verifySharing} function failed to trigger rehashing after copying shared hash structures, resulting in inconsistencies and an \texttt{ArrayIndexOutOfBoundsException} during repeated serialization. The fix introduces a new function, \texttt{\_verifyNeedForRehash}, to ensure proper rehashing and maintain hash table consistency.

As illustrated in Figure~\ref{fig:1} (a), although ChatGPT can reason about the need for rehashing and attempts to introduce the relevant logic, it lacks access to project-specific function names and implementation details. As a result, ChatGPT generates a plausible but incorrect function name, \texttt{rehashIfNecessary()}, based only on its general programming knowledge. In practice, however, the correct fix relies on a function, \texttt{\_verifyNeedForRehash}, that already exists within the project. This example highlights the critical importance of retrieving and providing relevant, project-specific information—such as existing function names and definitions—to the LLM. Supplying such context enables ChatGPT to generate precise and fully correct patches, effectively resolving the bug and restoring intended behavior in JacksonCore-11.

\subsection{Relevant Code Snippets Motivating Example}
\begin{figure}
    \centering
    \includegraphics[width=0.5\textwidth]{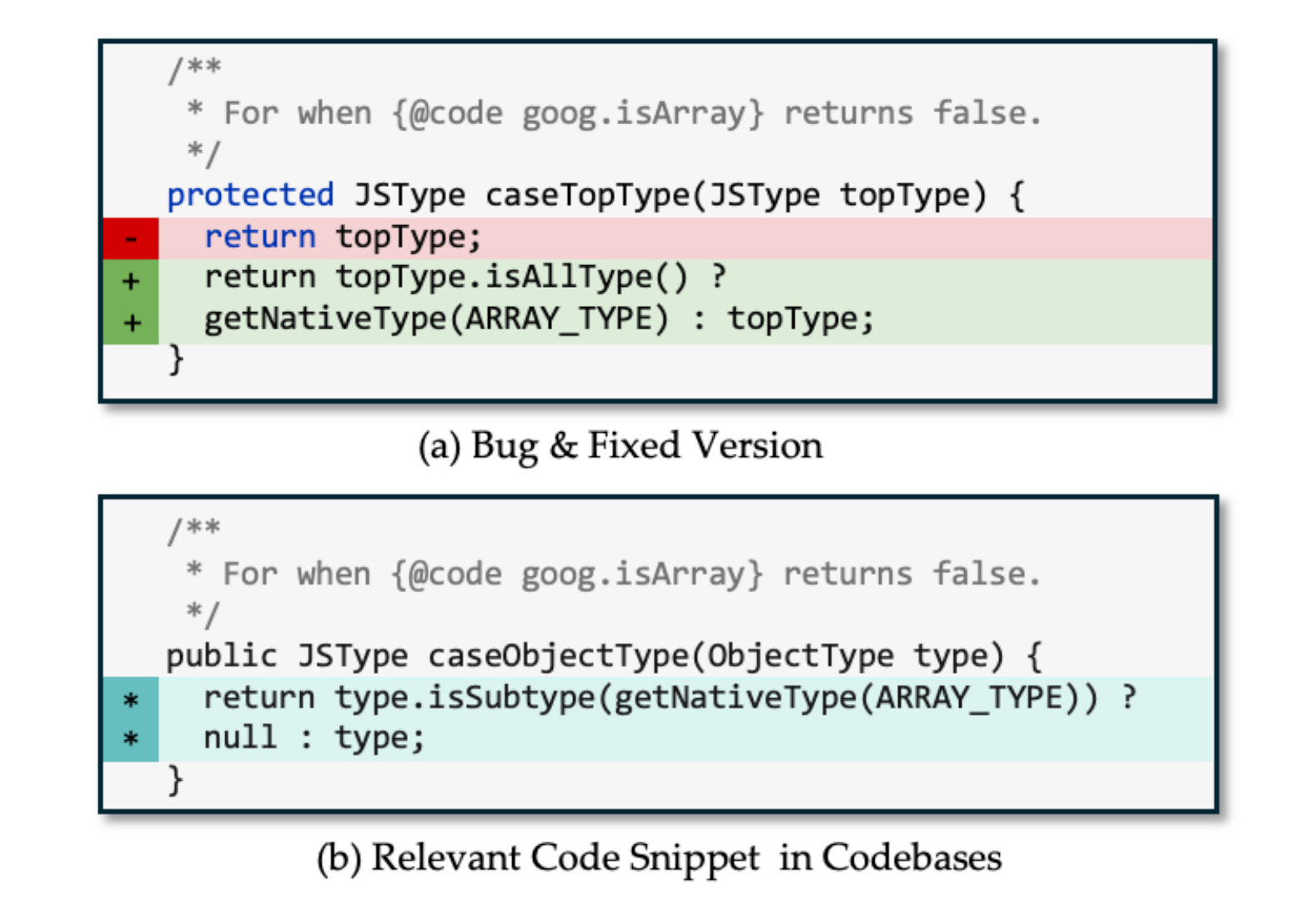}
    \caption{\centering Comparison of Buggy Version, Fixed Version and Relevant Code Snippet for \href{https://storage.googleapis.com/google-code-archive/v2/code.google.com/closure-compiler/issues/issue-1114.json}{Closure-111}}
    \label{fig:2}
\end{figure}

Figure~\ref{fig:2} presents the Closure-111 bug, illustrating a common challenge faced by LLM-based code repair. The bug in Figure~\ref{fig:2} (a) appears in the \texttt{caseTopType} function, which is responsible for classifying types in JavaScript Abstract Syntax Tree (AST) nodes. In the buggy version, the function incorrectly returns the input parameter \texttt{topType} without verifying whether it represents an array type. This omission results in misclassification and subsequent semantic errors in downstream components.

Repairing this bug is particularly challenging because it requires detailed, project-specific reasoning about the codebase’s type system—knowledge that is not available from generic programming data or typical LLM pretraining. The repair process demands an understanding of how array types are distinguished from other types and the specific conventions used in the Closure project for type handling.

This is precisely where \relrepair\ demonstrates its strength. Unlike generic retrieval-augmented methods that might surface broadly similar code or standard library examples, \relrepair\ is designed to retrieve code snippets that are highly relevant and specific to the project at hand. In this example, Figure~\ref{fig:2}(b) shows that \relrepair\ identifies the \texttt{caseObjectType} function as a top retrieval candidate. This function is not only located in the same project and file as the buggy code but also shares strong contextual similarities in both naming and documentation style.

By surfacing \texttt{caseObjectType}—a function that correctly distinguishes between different JavaScript types, including arrays—\relrepair\ provides the LLM with concrete, project-specific logic and implementation patterns that are directly applicable to fixing the bug. The LLM can then use this relevant context to infer the missing array-type check and synthesize a patch that aligns with the established type-handling semantics of the codebase. This targeted retrieval of high-value, project-specific information is a key enabler for effective and precise program repair, going beyond what generic RAG frameworks typically offer. 

\section{APPROACH}

We propose \relrepair, a novel retrieval-augmented APR approach designed to enhance LLM-based repair effectiveness. In contrast to existing LLM-based APR techniques, which typically generate patches exclusively from the provided buggy code, \relrepair systematically retrieves relevant information from the project's code repository. Specifically, it identifies and integrates function signatures (function name with their inputs) and related code snippets, thereby augmenting the input provided to the LLM. This targeted retrieval approach leverages project-specific knowledge to strengthen the LLM’s capacity for generating accurate and contextually appropriate patches. 

\begin{figure*}
    \centering
    \includegraphics[width=\textwidth]{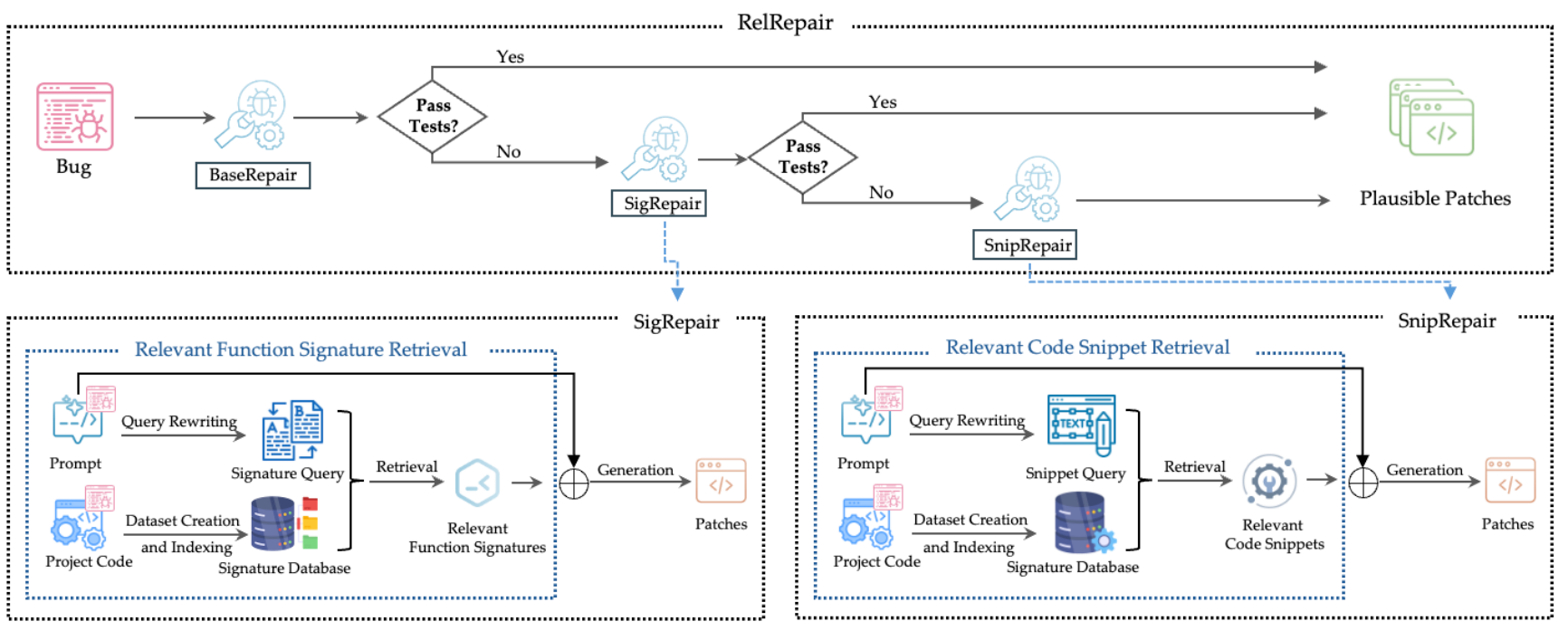}
    \caption{Overall Framework of RelRepair}
    \label{fig:overview}
\end{figure*}

Rather than using LLMs as standalone repair tools, \relrepair employs a RAG approach to systematically enhance the repair process. As depicted in Figure~\ref{fig:overview} and Algorithm ~\ref{alg:relrepair}, \relrepair consists of three incremental stages: \textbf{BaseRepair}, \textbf{SigRepair}, and \textbf{SnipRepair}. Initially, the BaseRepair stage attempts a straightforward repair using only the buggy function, error messages, failing test cases and perfect fault localization via LLM. This stage serves as an efficient first-pass filter to handle relatively simple bugs without relying on external information. If BaseRepair does not produce a correct patch, the process advances to the retrieval-based stages, which incorporate additional code to support the repair of more complex issues. If the patch generated at this stage fails to pass all tests, the repair process proceeds to the next stage, SigRepair. Likewise, if SigRepair is unsuccessful in producing a valid patch that passes the test suite, the process continues to the final stage, SnipRepair.

\begin{algorithm}[t]
\caption{\relrepair: Overall Algorithm of RelRepair}
\label{alg:relrepair}
\KwIn{Buggy function $f_{\text{bug}}$, failing test cases $\mathcal{T}_{\text{fail}}$, error messages $\mathcal{E}$, perfect fault localization $\mathcal{L}$}
\KwOut{Valid patch $p^*$ (if found) or $\emptyset$}

$p^* \gets \emptyset$\;

% --- BaseRepair Stage ---
\tcp*[h]{\textbf{Stage 1: BaseRepair}}\;
$\mathcal{P}_{\text{base}} \gets$ \textbf{BaseRepair}($f_{\text{bug}}, \mathcal{T}_{\text{fail}}, \mathcal{E}, \mathcal{L}$)\;
\ForEach{$p \in \mathcal{P}_{\text{base}}$}{
    \If{$p$ passes all test cases}{
        \Return $p $
    }
}

\tcp*[h]{\textbf{Stage 2: SigRepair}}\;
$\mathcal{P}_{\text{sig}} \gets$ \textbf{SigRepair}($f_{\text{bug}}, \mathcal{T}_{\text{fail}}, \mathcal{E}, \mathcal{L}$)\;
\ForEach{$p \in \mathcal{P}_{\text{sig}}$}{
    \If{$p$ passes all test cases}{
        \Return $p $
    }
}

\tcp*[h]{\textbf{Stage 3: SnipRepair}}\;
$\mathcal{P}_{\text{snip}} \gets$ \textbf{SnipRepair}($f_{\text{bug}}, \mathcal{T}_{\text{fail}}, \mathcal{E}, \mathcal{L}$)\;
\ForEach{$p \in \mathcal{P}_{\text{snip}}$}{
    \If{$p$ passes all test cases}{
        \Return $p $
    }
}

\Return $\emptyset$\
\end{algorithm}

The subsequent stages, SigRepair and SnipRepair, incorporate retrieved project-specific information into the LLM prompt, improving repair effectiveness. Specifically, SigRepair is a function signature-based repair method, utilizing \textit{Relevant Function Signature Retrieval} to augment the LLM input with signatures that effectively address simpler code modifications, such as function additions or replacements. After applying SigRepair, if a valid patch is generated that passes the test suite, we terminate the repair process for that bug and proceed to the next bug. If no valid patch is generated, the unresolved bug is passed to the next stage, SnipRepair. In contrast, SnipRepair leverages \textit{Relevant Code Snippet Retrieval} to integrate semantically similar code snippets into the LLM prompt, enabling it to more effectively address bugs that need project-specific information. The core components of \relrepair lies in its retrieval components, particularly the systematic retrieval of relevant function signatures and code snippets. Both retrieval processes consist of three structured phases: \blackcircled{1} \textbf{Query Rewriting}, optimizing queries based on the buggy function context; \blackcircled{2} \textbf{Dataset Creation and Indexing}, constructing customized datasets from the project codebase; and \blackcircled{3} \textbf{Retrieval}, selecting contextually relevant functions or code snippets to improve the accuracy of patches generated by LLM.

Using a structured and incrementally enhanced retrieval strategy, \relrepair improves the accuracy, relevance, and reliability of patches generated by LLM, enabling a more robust automated repair process. In the remainder of this section, we detail the design and methodology behind the \textit{Relevant Function Signature Retrieval} (Section~\ref{subsec:function_sig_retrieval}), \textit{Relevant Code Snippet Retrieval} (Section~\ref{subsec:code_snip_retrieval}), and their respective \textit{Generation} (Section~\ref{subsec:generation}) stages.

\subsection{Relevant Function Signature Retrieval}  \label{subsec:function_sig_retrieval}

\subsubsection{Query Rewriting} \label{subsubsec: query_rewriting_sig}
Our basic prompt—"\textsf{\texttt{You are an expert in program repair.}}"—supplied to ChatGPT along with the buggy function, failing test, error messages and perfect fault localization, is sufficient for an initial patch attempt but inadequate for precisely retrieving relevant function signatures. In particular, SigRepair typically relies on inserting or replacing critical function calls or identifiers. However, directly using the buggy function as a retrieval query often yields irrelevant matches due to limited textual similarity with the correct functions. To overcome this, we leverage ChatGPT's reasoning capabilities by prompting it ($prompt_{SIG}$) to generate two root causes and five candidate function names likely to repair the buggy line. Limiting the number to two root causes and five function names avoids semantic redundancy while capturing plausible alternatives. By explicitly rewriting retrieval queries in this targeted manner, we substantially increase the likelihood of retrieving the most relevant function signatures. Both this section and the subsequent procedure are detailed in Algorithm~\ref{alg:sigrepair}.

\begin{algorithm}[t]
\caption{SigRepair: Function Signature-based Repair}
\label{alg:sigrepair}
\KwIn{Buggy function $f_{\text{bug}}$, failing test cases $\mathcal{T}_{\text{fail}}$, error messages $\mathcal{E}$, perfect fault localization $\mathcal{L}$}
\KwOut{Set of candidate patches $\mathcal{P}_{\text{sig}}$}

$\mathcal{P}_{\text{sig}} \gets \emptyset$ 

\For{$i \gets 1$ \KwTo $20$}{
 \tcp*[h]{\textbf{Query Rewriting}}
 
  $q \gets$ LLM($prompt_{SIG}$, $f_{\text{bug}}$)\;
  $\mathbf{e}_{\text{qry}} \gets $ SentenceBERT$(q)$\;

  \tcp*[h]{\textbf{Dataset Creation and Indexing}}

  $\mathcal{D}_{\text{sig}} \gets \texttt{File}(f_{\text{bug}}) \cup \texttt{Type}(f_{\text{bug}})$\; 

  \ForEach{$f \in \mathcal{D}_{\text{sig}}$}{
  $\mathbf{e}_f \gets \mathrm{SBERT}(\text{sig}(f) \parallel \text{cmt}(f))$\;
    \tcp*[h]{\textbf{Retrieval}}
  $\mathrm{sim}(\mathbf{e}_{\text{qry}}, \mathbf{e}_i) = 
    \frac{\mathbf{e}_{\text{qry}} \cdot \mathbf{e}_i}
     {\lVert \mathbf{e}_{\text{qry}} \rVert \, \lVert \mathbf{e}_i \rVert}
    \quad \forall \, \mathbf{e}_i \in \mathbf{E}_{\text{cand}}$ $\mathbf{e}_f$\;
    
  }
  $\mathcal{F}_{\text{top}} \gets \operatorname{Top}_{25}(\mathrm{sim}(\mathbf{e}_{\text{qry}}, \mathbf{e}_i)),\ \forall \mathbf{e}_i \in \mathbf{E}_{\text{cand}}$\;

  \tcp*[h]{\textbf{Patch Generation}}
   $\text{prompt} \gets f_{\text{bug}}, \mathcal{T}_{\text{fail}}, \mathcal{E}, \mathcal{L}, \text{and } \mathcal{F}_{\text{top}}$\;
  $p_i \gets$ LLM$(\text{prompt},\ \text{num\_return} = 1))$\;
  $\mathcal{P}_{\text{sig}} \gets \mathcal{P}_{\text{sig}} \cup \{p_i\}$\;
}

\Return $\mathcal{P}_{\text{sig}}$\;
\end{algorithm}

\subsubsection{Dataset Creation and Indexing}
To enhance the effectiveness of bug repair, we construct a tailored dataset for each individual bug, enabling the precise retrieval of relevant function signatures and effectively narrowing the retrieval scope. This strategy ensures that the retrieved functions are specifically pertinent to addressing the targeted bug, thereby improving the quality and accuracy of generated fixes. Our selection strategy leverages the observation that fixes commonly involve functions that either (1) reside in the same file as the buggy function or (2) relate to author-defined variable types explicitly introduced within the buggy function. The methodology for systematically identifying these relevant functions is detailed as follows:

We define the set of relevant functions as a combination of two main parts: variable-based functions and file-based functions. Variable-based functions refer to functions associated with custom-defined types used within the buggy function. Each variable type defined by the original author typically comes with a set of related functions. For example, if a variable type represents a particular data structure, there are typically multiple functions that operate on or interact with that structure. File-based functions include functions who exist within the same source code file as the buggy function. This ensures that functions closely related by context or functionality are considered relevant.
By combining these two sets, we obtain a highly targeted pool of candidate functions. This effectively narrows down the retrieval scope, making the selected functions more directly related and useful for repairing the specific bug.

For every function in the dataset, we extract two important details: (1) Function Signature: This includes the name of the function and the parameters it accepts, giving a clear snapshot of its interface. (2) Block Comments: These comments provide essential context and concise explanations about the function's intended behavior, usage scenarios, and underlying purpose. To identify candidate functions for SigRepair, we encode both the function signature and its corresponding block comment using SentenceBERT, resulting in dense vector embeddings for each entry in our retrieval dataset. This decomposition enables effective retrieval by facilitating comparisons between block comments and root cause descriptions, as well as matching function signatures with candidate function names, thereby improving both the accuracy and effectiveness of the retrieval process.

\subsubsection{Retrieval}
 During query rewriting (\ref{subsubsec: query_rewriting_sig}) LLM produces (i) a concise root cause description and
(ii) up to five plausible function names. We concatenate these two elements, encode the resulting text with SentenceBERT, and obtain a single query embedding.
Retrieval is then performed using cosine similarity. Subsequently, we select the top \(K=25\) functions exhibiting the highest similarity scores. Each of these functions, represented by its signature and corresponding comment, is integrated into the LLM prompt, providing targeted contextual information to facilitate patch generation.

\subsection{Relevant Code Snippet Retrieval} \label{subsec:code_snip_retrieval}
SnipRepair assumes the existence of relevant functions that are structurally and semantically similar to the buggy function provided within the project. Such functions typically contain logical constructs or function usages directly applicable to fixing the identified bug. Leveraging this assumption, SnipRepair first identifies these relevant functions based on structural similarity and functional intent. Then, guided by the provided precise fault localization of the buggy function, SnipRepair selectively extracts code snippets from these relevant functions. This process is detailed in Algorithm~\ref{alg:sniprepair}.

\begin{algorithm}[t]
\caption{SnipRepair: Code Snippet-based Repair}
\label{alg:sniprepair}
\KwIn{Buggy function $f_{\text{bug}}$, failing test cases $\mathcal{T}_{\text{fail}}$, error messages $\mathcal{E}$, perfect fault localization$\mathcal{L}$}
\KwOut{Set of candidate patches $\mathcal{P}_{\text{snip}}$}

$\mathcal{P}_{\text{snip}} \gets \emptyset$ 

\tcp*[h]{\textbf{Query Rewriting}}

Separate $f_{\text{bug}}$ into code and comment components\;
$E_{qf} \gets \text{CodeBERT}(\text{code}(f_{\text{bug}}))$\;
$E_{qc} \gets \text{CodeBERT}(\text{comment}(f_{\text{bug}}))$\;

\tcp*[h]{\textbf{Dataset Creation and Indexing}}

    $\mathcal{F}_{\text{intra}} \gets$ functions in same file as $f_{\text{bug}}$\;
    $\mathcal{F}_{\text{inter}} \gets$ functions from top-5 similar files in same directory\;
    $\mathcal{D}_{\text{snip}} \gets \mathcal{F}_{\text{intra}} \cup \mathcal{F}_{\text{inter}}$\;
    
    \ForEach{$f \in \mathcal{D}_{\text{snip}}$}{
      $E_f \gets \text{CodeBERT}(\text{code}(f))$\;
      $E_c \gets \text{CodeBERT}(\text{comment}(f))$\;
      \tcp*[h]{\textbf{Retrieval}}
      $\text{Sim}(f) \gets \alpha \cdot \frac{E_{qf} \cdot E_f}{\|E_{qf}\| \|E_f\|} + 
                          \beta \cdot \frac{E_{qc} \cdot E_c}{\|E_{qc}\| \|E_c\|}$\;
    }
    
    $\mathcal{F}_{\text{top}} \gets \operatorname{Top}_{30}(\text{Sim}(f))$\;
    $\mathcal{F}_{\text{top}} \gets \text{Sort}(\mathcal{F}_{\text{top}}, \downarrow \text{Sim})$\;

    \tcp*[h]{\textbf{Patch Generation}}
    
    \ForEach{$f_k \in \mathcal{F}_{\text{top}}$}{
      $\text{prompt} \gets \{f_{\text{bug}}, \mathcal{T}_{\text{fail}}, \mathcal{E}, \mathcal{L}, f_k\}$\;
      $\mathcal{P}_k \gets$ LLM$(\text{prompt},\ \text{num\_return} = 10)$\;
      \ForEach{ $i \in 1\ldots 10$} {
       \If{$p_{k,i}$ passes all tests}{
          \Return $\mathcal{P}_{\text{snip}} \cup \{p_{k,i}\}$\;
        }
        $\mathcal{P}_{\text{snip}} \gets \mathcal{P}_{\text{snip}} \cup \{p_{k,i}\}$\;
      }    
    }
    
    \Return $\mathcal{P}_{\text{snip}}$\;
\end{algorithm}

\subsubsection{Query Rewriting}
Rewriting of code snippets focuses on extracting meaningful code snippets that align closely with the fix. We preprocess the buggy function by separating inline and block comments from the code, maintaining these components separately. By doing so, we create independent query components, including the buggy function and its associated comments, which allows for targeted retrieval of semantically similar code snippets. By isolating these elements, SnipRepair ensures that each retrieval query precisely targets behaviorally relevant code snippets, thus improving retrieval precision for downstream patch generation.

\subsubsection{Dataset Creation and Indexing}
Code snippet dataset construction involves selecting the most relevant code snippets from large-scale Java projects. First, we collect all functions located within the same file as the buggy function (intra-file), as these are likely to share contextual dependencies. Second, we identify the five most signature-similar files within the same directory and extract all functions defined in those files (inter-file), reflecting broader but still project-specific context. For each extracted function, we further separate the function body from its accompanying inline and block comments to facilitate structured processing. We utilize the pre-trained model, CodeBERT, which is explicitly trained to understand the semantics of source code, enabling us to derive robust semantic representations. For each candidate function $f_i$ and its associated comments $c_i$, we generate separate dense embeddings using CodeBERT.

\subsubsection{Retrieval}
To conduct retrieval, we leverage a strategy similar to the function signature retrieval approach but specifically tailored towards capturing deeper semantic relationships within the source code. Unlike SigRepair, which primarily considers surface-level features such as function signatures and descriptive comments, our SnipRepair emphasizes the internal logic and structural semantics inherent within code snippets.

To quantify the semantic similarity between the rewritten query and the candidate code snippets, we independently assess the resemblance of both the function code and its associated comments. Specifically, we use CodeBERT to generate embeddings for each component, and then evaluate how closely the query’s function embedding matches that of the candidate, as well as how well the query’s comment embedding aligns with the candidate’s comment embedding. Each similarity score is computed separately for the function code and for the descriptive comments, thereby capturing both the structural and semantic relationships between the query and the candidate. Additionally, we introduce two weighting factors, $\alpha$ and $\beta$, which control the relative importance of the function embeddings and the comment embeddings, respectively. This approach enables a balanced consideration of both structural and descriptive aspects.

To determine appropriate values for these weights, we do not rely solely on heuristic assumptions. Instead, we employ a lightweight, iterative gradient-based adjustment method. The weights are initialized equally, with $\alpha = 0.5$ and $\beta = 0.5$. During each iteration, the weighted similarity is computed as a linear combination of the structural similarity (function embeddings) and semantic similarity (comment embeddings), and $\alpha$ and $\beta$ are adaptively adjusted based on their contributions toward minimizing the gap between the current similarity and a predefined target similarity. As a result, if structural similarity is found to play a more substantial role in approaching the target, $\alpha$ gradually increases while $\beta$ correspondingly decreases, and vice versa. 

Finally, we select the top $K=15$ most semantically relevant candidate functions from both \emph{intra-file} and \emph{inter-file} contexts, resulting in a total of 30 code snippets. Specifically, the 15 highest-scoring candidates are chosen separately from within the same file as the query (intra-file) and from other files in the project (inter-file). 

\subsection{Generation} \label{subsec:generation}
In \textit{SigRepair}, we retrieve the top 25 function signatures that are most relevant to the buggy function, based on similarity. These signatures are collectively injected into the LLM prompt to guide patch generation. Since the effectiveness of this stage depends on the initial query rewriting stage, we repeat the \textit{Relevant Function Signature Retrieval} process 20 times, each time producing a single candidate patch. As a result, the SigRepair stage yields up to 20 candidate patches. 

In \textit{SnipRepair}, we retrieve 15 semantically relevant code snippets from the intra-file context and another 15 from the inter-file context, with each group ranked independently by their similarity to the buggy function. During patch generation, we process all intra-file snippets first. For each snippet, we generate 10 candidate patches and immediately validate them using the test suite; if all patches fail, we proceed to the next snippet. Only after all intra-file snippets have been exhausted do we process the inter-file snippets in the same manner. This strategy reflects our experimental finding that the most relevant code is often contained within the same file as the buggy function. While SnipRepair can theoretically produce up to 300 patches (30 snippets × 10 patches), the actual number is typically lower due to early stopping once a successful patch is found.

\section{EXPERIMENTAL DESIGN}
We evaluate \relrepair on the following research questions:
\begin{itemize}
    \item RQ1: How does the performance of \relrepair compare with the state-of-the-art LLM-based APRs?
    \item RQ2:  How do different LLMs influence the effectiveness of \relrepair?
    \item RQ3:  How do different components of \relrepair affect the overall performance of \relrepair?
\end{itemize}

\subsection{Datasets}

\begin{table*}[ht]
\centering
\caption{Correct Fixes (Single-Function Setting) across Repair Tools on Defects4J V1.2}
\setlength{\tabcolsep}{6pt} 

\begin{tabular}{
    >{\centering\arraybackslash}p{1.2cm} % Project
    |>{\centering\arraybackslash}p{1.8cm} % RelRepair
    |>{\centering\arraybackslash}p{1.8cm} % BaseChatGPT
    |>{\centering\arraybackslash}p{1.8cm} % ThinkRepair
    |>{\centering\arraybackslash}p{1.8cm} % Muplor
    |>{\centering\arraybackslash}p{1.8cm} % ChatRepair
    |>{\centering\arraybackslash}p{1.8cm} % RAP-Gen
}
\toprule
\textbf{Project} 
& \textbf{RelRepair} 
& \textbf{BaseChatGPT}
& \textbf{ThinkRepair}
& \textbf{Muplor}
& \textbf{ChatRepair}
& \textbf{RAP-Gen} \\
\midrule
Chart   & 10 & 5  & \ 11 & 10 & / & 7  \\
Closure & 35 & 10 & 31 & 22 & / & 16 \\
Lang    & 19 & 8  &  19 & 14 & / & 7  \\
Math    & 28 & 18 & 27 & 17 & / & 10 \\
Mockito & 5  & 4  & 5  & 5  & / & 1  \\
Time    & 4  & 2  & 2  & 2  & / & 1  \\
\midrule
\textbf{\#Sum} & 101 & 47 & 98 & 70 & 76 & 42 \\
\bottomrule
\end{tabular}
\label{tab:defects4j_results}
\end{table*}

\setlength{\tabcolsep}{4pt} % Reduce column spacing
\begin{table}[h]
    \centering
    \footnotesize
    \caption{Statistics of studied dataset}
    \label{tab:dataset_statistics}
    \begin{tabular}{lrrrr}
        \toprule
        \textbf{Dataset} & {\textbf{\# Total Bugs}} & {\textbf{\# SF Bugs}} & {\textbf{\# SH Bugs}} & {\textbf{\# SL Bugs}} \\
        \midrule
        Defects4J 1.2 & 391 & 255 & 154 & 80 \\
        ManySStuBs4J & 480 & 480 & 480 & 480 \\
        \midrule
        \textbf{\# Sum} & \textbf{871} & \textbf{735} & \textbf{634} & \textbf{560} \\
        \bottomrule
    \end{tabular}
    \vspace{1mm}
\end{table}

For the evaluation, we follow previous studies to adopt two widely used APR benchmarks: Defects4J V1.2 \cite{just2014defects4j} and ManySStuBs4J \cite{karampatsis2020often}. While previous research often includes the QuixBugs \cite{lin2017quixbugs} dataset, many LLM-based APRs have demonstrated strong performance on QuixBugs, making it less challenging for evaluation. Therefore, we select Defects4J V1.2 and ManySStuBs4J to ensure a more rigorous assessment. Following the previous APR studies \cite{xia2023automated, xia2022less, xia2023keep}, we use Defects4J 1.2 which contains 391 bugs across six Java projects. We focus on single-function bugs, as this setting is commonly adopted in most recent studies \cite{jiang2021cure}, \cite{lutellier2020coconut}, \cite{xia2023automated}. Additionally, we include ManySStuBs4J, a large-scale dataset comprising over 150,000 single-statement Java bugs, categorized into 16 distinct bug patterns. The dataset provides detailed commit hashes and repository identifiers for each instance, along with both the buggy and fixed versions of the code, enabling a fine-grained evaluation of APR techniques. However, due to the extensive size of ManySStuBs4J, evaluating all instances would require significant computational resources and time. To balance efficiency and representativeness, we select 30 bugs from each category, resulting in a total of 480 bugs for evaluation.

\subsection{Baselines and Metrics}
We evaluate \relrepair against four state-of-the-art LLM-based APR techniques — ChatRepair \cite{xia2023keep}, ThinkRepair \cite{yin2024thinkrepair}, Mulpor \cite{lin2024one} and RAP-Gen \cite{wang2023rap}, as well as a baseline method, BaseChatGPT, which performs code repair by directly sampling from the ChatGPT model without access to any retrieved relevant code. To ensure a comprehensive comparison, we adopt this widely used evaluation metrics: correct patches, which are patches that pass all the test cases and are semantically equivalent to the actual fix. Following prior work \cite{xia2023automated, xia2022less}, we also conduct manual validation to identify plausible patches that are functionally equivalent to the actual fixes. For baseline results, we reference reported outcomes from prior studies where available, ensuring a consistent and fair comparison across all approaches.

\subsection{Implementation}
We implement the core logic of \relrepair in Python by leveraging the gpt-3.5 model from the ChatGPT family, ensuring fair comparison with prior APR studies that also utilize similar LLM-based configurations\cite{yin2024thinkrepair}, \cite{xia2023keep}. To keep a diverse set of potential patches, we set the sampling temperature to 1 \cite{gilardi2023chatgpt}. For the BaseRepair process, we generate a single candidate patch. During the SigRepair stage, we expand this to 20 candidate patches, while in the SnipRepair stage, we further increase the number to 300 candidates. Consistent with prior APR work \cite{xia2022less, zhu2021syntax}, we set a default end-to-end timeout of 5 hours for fixing a single bug. However, in practice, the average repair time is significantly lower, typically under 20 minutes, as we employ a small-scale patch sampling strategy for each bug, reducing computational overhead.

\section{EVALUATION}
\subsection{RQ1: LLM-based APRs Comparision}
\subsubsection{\textbf{Experimental Design}}
Our RelRepair approach leverages retrieval-augmented patch generation based on two prominent datasets: Defects4J V1.2 and ManySStuBs4J.

\textbf{Defects4J V1.2}. As all studied baselines have been previously evaluated on Defects4J V1.2, we directly compare RelRepair's automated repair results against these existing studies under identical experimental settings. Specifically, we examine three widely recognized repair scenarios commonly used in prior APR research \cite{xia2023automated, xia2023keep}: single-function (SF), single-hunk (SH), and single-line (SL). Notably, single-hunk are a subset of single-function, and single-line are a subset of single-hunk bugs. All experiments are conducted under the single-function repair setting, which is the primary setting adopted by most prior APR tools\cite{xia2023automated, xia2023keep, yin2024thinkrepair}, to ensure fair comparison. For evaluations involving BaseChatGPT on Defects4J V1.2, we provide both the test cases and error messages available in the dataset to facilitate comprehensive analysis by ChatGPT.

\textbf{ManySStuBs4J}. ManySStuBs4J encompasses 16 distinct bug categories drawn from real-world Java Maven projects. Unlike Defects4J V1.2, designed explicitly to support reproducible research scenarios, ManySStuBs4J provides insights into practical software defects frequently encountered in open-source environments. The ManySStuBs4J dataset exclusively contains single-line bug fixes, and hence, our experiments for this dataset are restricted to the single-line scenario. For a thorough assessment, we select 30 bugs from each category within ManySStuBs4J. As the dataset lacks test suites, we cannot provide failing test cases and error messages during patch generation. Consequently, the correctness of generated fixes is assessed solely by directly comparing the generated fix to the provided ground-truth solution; only exact matches are considered correct. We reconstruct complete project versions using commit information from the dataset to ensure accurate contextual analysis. 

\textbf{Fault Information.} Following conventions established in previous APR research \cite{jiang2021cure, lutellier2020coconut, xia2022less, xia2023automated, ye2022neural}, we assume perfect fault information (FL), including detailed statement-level fault information, to maintain consistency across comparative evaluations.

\subsubsection{\textbf{Results}}
\textbf{Defects4J V1.2.} Table \ref{tab:defects4j_results} illustrates the number of bugs successfully repaired by \relrepair and other LLM-based APR approaches as well as BaseChatGPT—across three scenarios in the Defects4J V1.2 benchmark. As summarized in Table \ref{tab:defects4j_results}, \relrepair achieves the highest number of successful repairs, fixing a total of 101 bugs. This represents a substantial improvement over BaseChatGPT, which repaired only 47 bugs, amounting to a performance increase of approximately 112.8\%.

\relrepair demonstrates substantial improvements over BaseChatGPT, particularly in challenging projects such as Closure, where it fixes 35 bugs compared to BaseChatGPT’s 10 - a 250\% increase. Similar gains are observed in Math, with \relrepair resolving 35\% more bugs. Furthermore, \relrepair exceeds other leading LLM-based APR approaches, repairing more bugs; for example, it achieves 38.9\% more fixes than RAP-Gen. It is important to note that our evaluation focuses on comparisons with LLM-based APR models, as competitive baselines like ThinkRepair have already demonstrated clear superiority over traditional NMT-based APR methods. Consequently, direct comparisons with NMT-based approaches are omitted, given that existing LLM-based methods have established their advantage in prior work.

\begin{figure}
    \centering
    \includegraphics[width=0.45\textwidth]{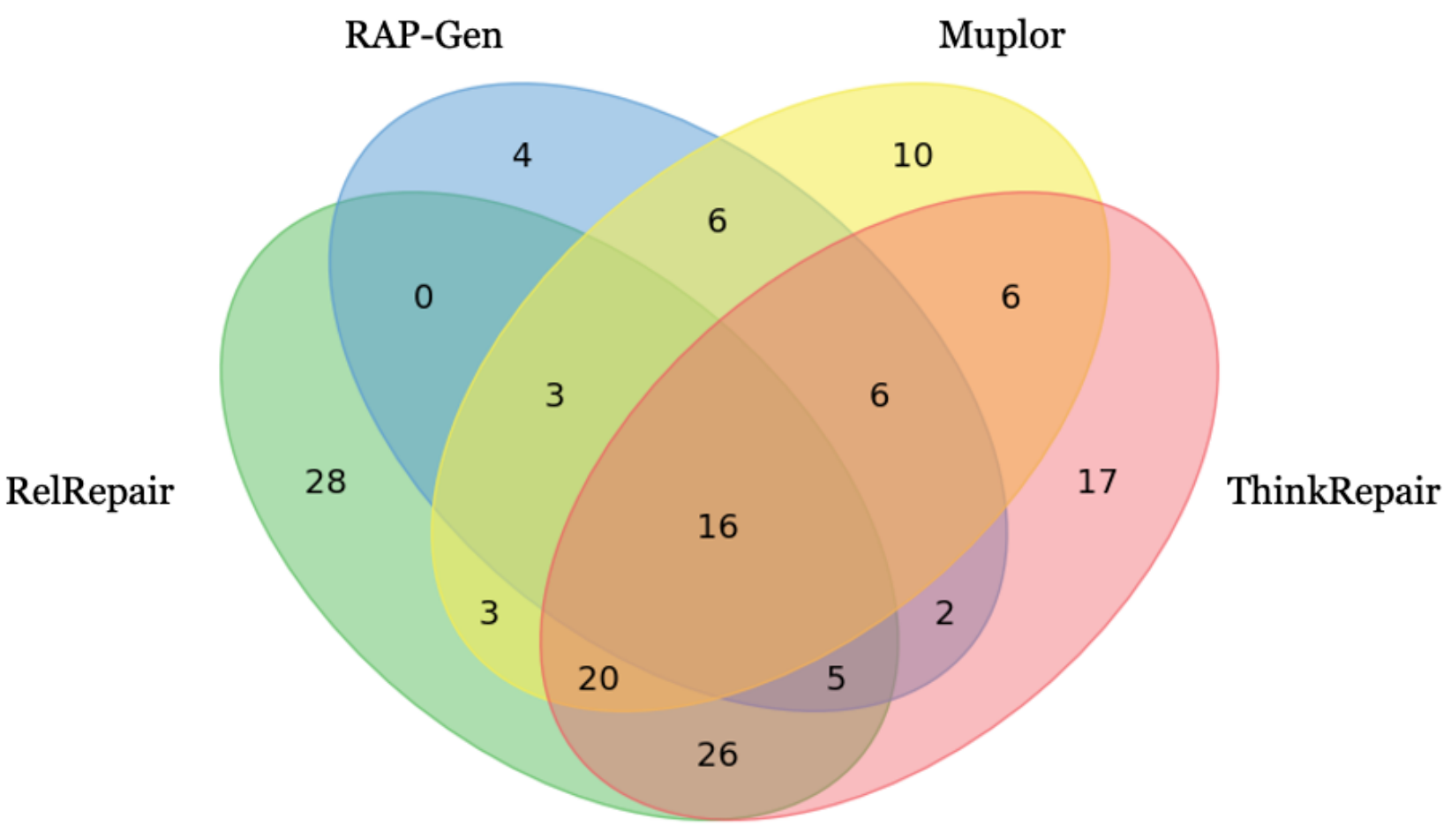}
    \caption{Venn diagram on Defects4J 1.2}
    \label{fig:venn}
\end{figure}

Fig.~\ref{fig:venn} presents a Venn diagram comparing the bugs fixed by \relrepair, RAP-Gen, Mulpor, and ThinkRepair on the Defects4J V1.2 dataset. We exclude ChatRepair from this comparison due to the unavailability of its generated patches on GitHub. Two key observations emerge from this comparison:
\textbf{(1)} In terms of total repair effectiveness, \relrepair achieves the highest number of fixed bugs, resolving \textbf{101} in total. While this number is only slightly higher than that achieved by ThinkRepair, the improvement over other state-of-the-art APR methods is substantial.
\textbf{(2)} More notably, \relrepair uniquely fixes \textbf{28} bugs that are not addressed by any of the other three state-of-the-art APR systems, the highest number of uniquely fixed bugs among all methods.

Despite their comparable total fix rates, the fundamental difference between these two LLM-based APRs lies in their underlying methodologies. Specifically, \relrepair employs a RAG strategy, retrieving relevant function signatures and code snippets from the codebase in which the buggy function resides to effectively guide the LLM. Conversely, ThinkRepair relies primarily on CoT prompting, guiding LLM through deeper logical reasoning processes without significant dependence on project-specific knowledge. Consequently, \relrepair excels in scenarios involving built-in functions or logic, whereas ThinkRepair demonstrates superior performance in cases requiring extensive reasoning capabilities. Moreover, the results highlight a substantial degree of complementarity between \relrepair and other approaches based on supervised learning, suggesting that the integration of these distinct methodologies could potentially produce enhanced overall bug-repair performance.

To demonstrate the effectiveness of \relrepair over BaseChatGPT, we examine a bug from the Defects4J V1.2 dataset (\texttt{Chart-10}) that is successfully fixed only by \relrepair. The bug involves improper handling of special characters, such as quotation marks, in HTML tooltip fragments, resulting in malformed HTML during testing. BaseChatGPT's attempted fix fails to escape these characters, leading to incorrect output and even compilation errors. In contrast, \relrepair retrieves a semantically relevant code snippet from the project codebase and correctly applies the \texttt{ImageMapUtilities.htmlEscape()} method. This approach ensures proper escape of special characters and generates a correct and compilable patch. 

\begin{figure}
    \centering
    \includegraphics[width=0.45\textwidth]{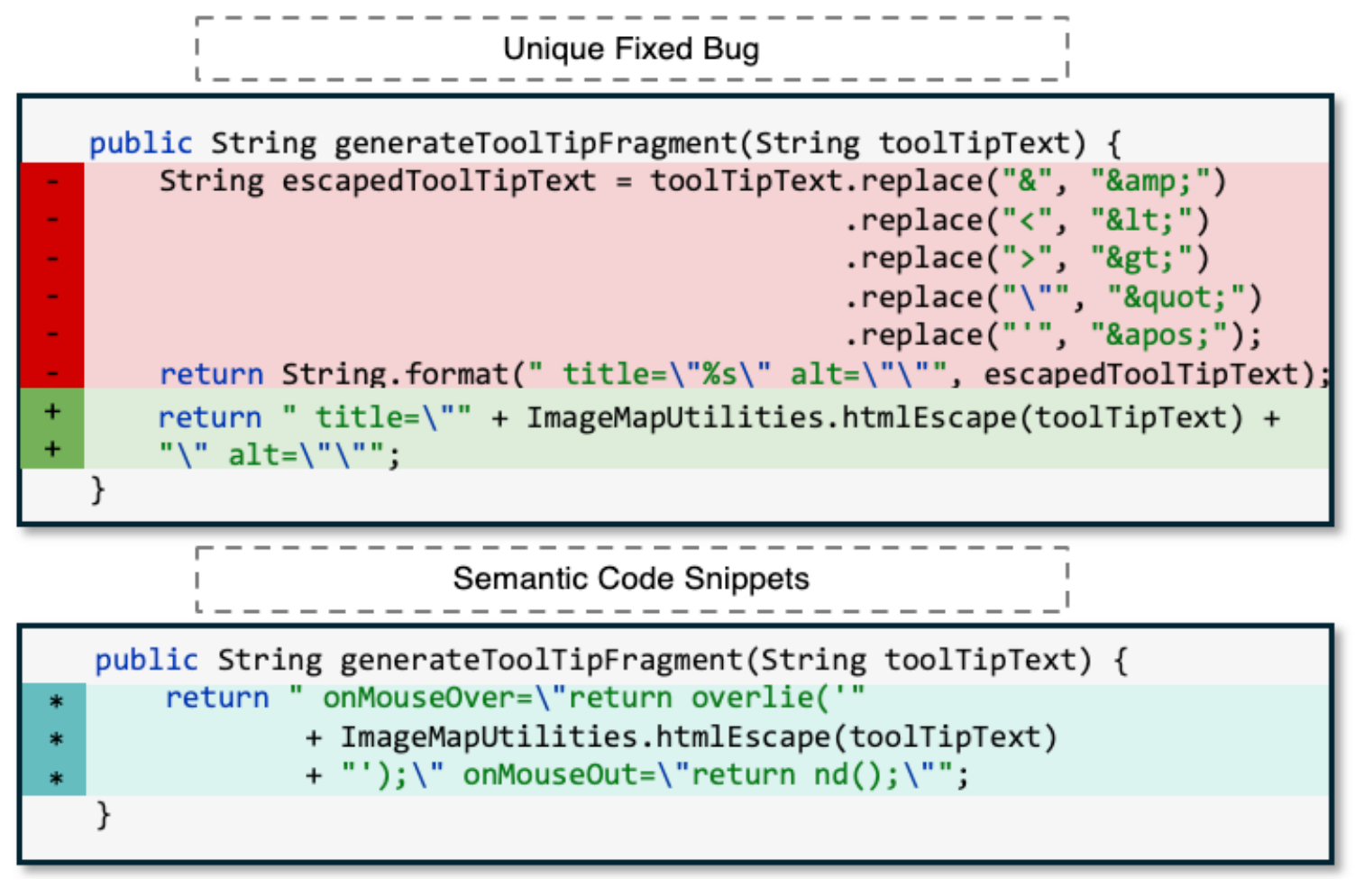}
    \caption{Unique bug fixed by \relrepair on Defects4J V1.2}
    \label{fig:rq1}
\end{figure}

\begin{figure}
  \centering
  \includegraphics[width=0.45\textwidth]{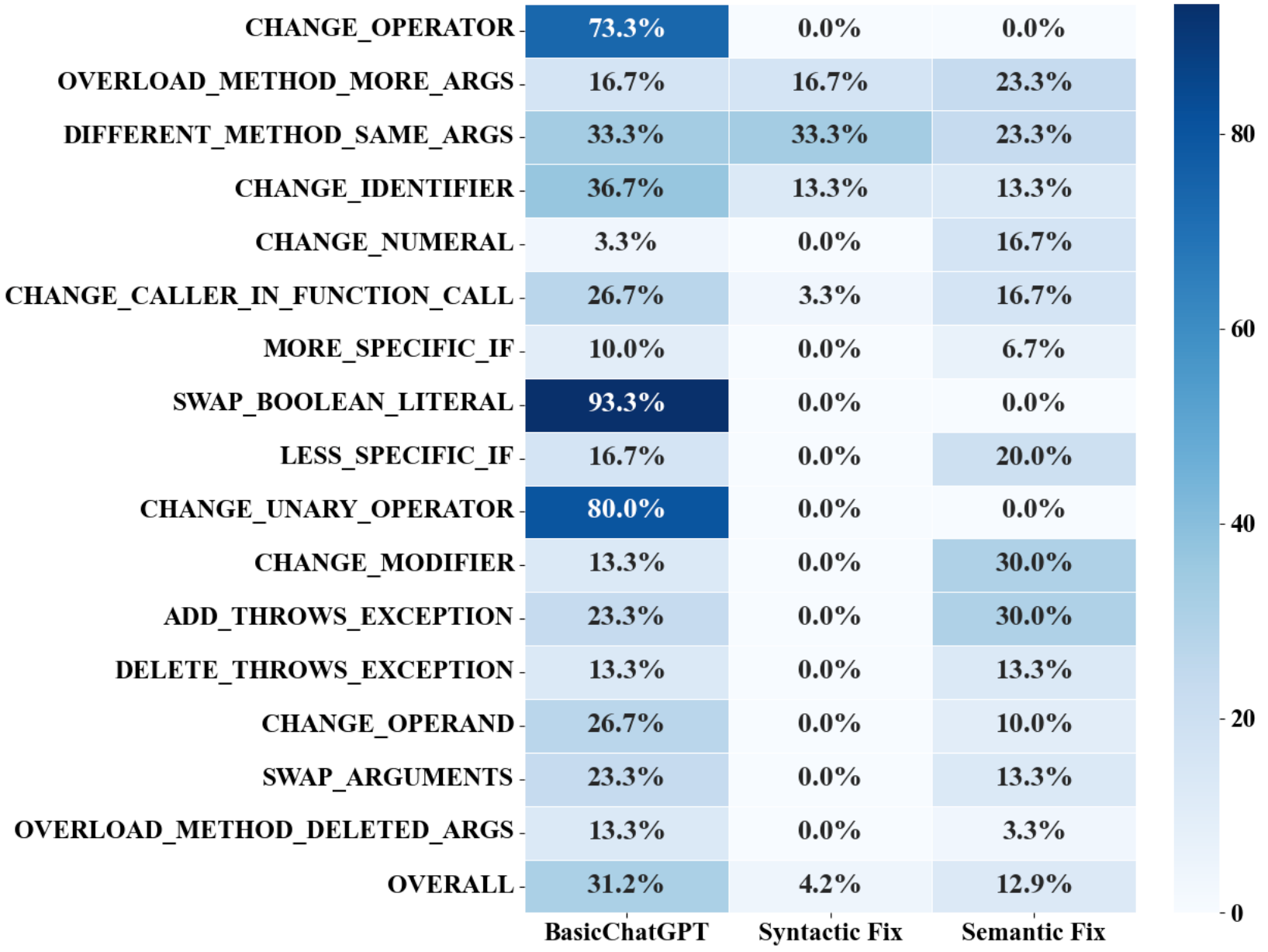}
  \caption{Fix Rate for ManySStuBs4J}
  \label{fig:manyststubs4j}
\end{figure}

\begin{table*}[ht]
    \centering
    \caption{RQ2: RelRepair vs. Basic LLMs for different projects on Defects4J V1.2}
    \resizebox{\textwidth}{!}{ 
    \begin{tabular}{l|cccc|cccc}
    \toprule
    \multirow{2}{*}{Projects} & \multicolumn{4}{c|}{LLMs with ReIRrepair} & \multicolumn{4}{c}{Basic LLMs} \\
    \cmidrule(lr){2-5} \cmidrule(lr){6-9}
                              & ChatGPT-4 & ChatGPT-3.5 & CodeLlama-13b & DeepSeek-33b & BaseChatGPT-4 & BaseChatGPT-3.5 & BaseCodeLlama-13b & BaseDeepSeek-33b \\
    \midrule
    Chart    & \cellcolor{gray!40}10 & \cellcolor{gray!40}10 & 6  & 7  & 7  & 5  & 4  & 6 \\
    Closure  & \cellcolor{gray!40}45 & 35 & 17 & 22 & 23 & 10 & 7  & 5 \\
    Lang     & \cellcolor{gray!40}23 & 19 & 12 & 14 & 20 & 8  & 7  & 8 \\
    Math     & \cellcolor{gray!40}31 & 28 & 19 & 17 & 28 & 18 & 9  & 10 \\
    Mockito  & \cellcolor{gray!40}6  & 5  & 5  & 5  & 4  & 4  & 4  & 5 \\
    Time     & \cellcolor{gray!40}5  & 4  & 2  & 3  & 4  & 2  & 1  & 0 \\
    \midrule
    \textbf{\# Sum} & \cellcolor{gray!40}\textbf{120} & \textbf{101} & \textbf{59} & \textbf{68} & \textbf{86} & \textbf{47} & \textbf{32} & \textbf{34} \\
    \bottomrule
    \end{tabular}
    \label{tab:RelRrepair vs. Basic LLMs}
    }
\end{table*}

\textbf{ManySStuBs4J}. We first evaluate the effectiveness of \relrepair using the ManySStuBs4J dataset, focusing on 16 distinct bug categories extracted from real-world Java Maven projects. Table \ref{fig:manyststubs4j} summarizes repair success rates by \relrepair under three fix strategies: BaseRepair (without relevant information), SigRepair, and SnipRepair. The reported value for each category represents the percentage of bugs fixed within that category—specifically, the fraction of successful repairs out of 30, expressed as a percentage. The final row, labeled “Overall,” reports the aggregate success rate across all categories, calculated as the total number of bugs fixed divided by the overall number of bugs in the dataset (480), giving an overall repair rate for each strategy.

In general, \relrepair achieves an improved total fix rate of 48.3\%, significantly exceeding the BaseRepair approach, which resolves 31.2\% of the bugs. In particular, the integration of relevant information markedly enhances performance, adding an additional 83 bug fixes beyond those achieved by the BaseRepair alone. Detailed examination reveals that certain bug categories benefit substantially from the retrieval-augmented strategy. For example, the categories DIFFERENT\_METHOD\_SAME\_ARGS, OVERLOAD\_METHOD\_MORE\_ARGS, CHANGE\_MODIFIER, and ADD\_THROWS\_EXCEPTION observe considerable improvements when relevant functions or relevant code snippets are retrieved, resulting in cumulative fixes of 15, 9, 9, and 9 bugs respectively. Moreover, the fix with the SnipRepair strategy significantly improves categories such as CHANGE\_NUMERAL, CHANGE\_CALLER\_IN\_FUNCTION\_CALL, and LESS\_SPECIFIC\_IF, demonstrating the necessity and effectiveness of retrieving contextually relevant code snippets.

\begin{tcolorbox}[colback=white, colframe=black, arc=5pt, boxrule=0.8pt, left=2pt, right=2pt, top=2pt, bottom=2pt]
    \textbf{Findings to \textbf{RQ1}:} 
\textit{Basic LLMs exhibit limited effectiveness in fixing bugs. In contrast, RelRepair successfully repairs \textbf{101} bugs, outperforming existing LLM-based APR approaches. It also resolves \textbf{28} bugs uniquely, which are not fixed by any other baseline. This improved capability highlights RelRepair's strength in effectively handling more complex and challenging bug scenarios.}
\end{tcolorbox}

\subsection{RQ2: Different LLMs on the Effectiveness of \relrepair }

\subsubsection{\textbf{Experimental Design}}
We establish baseline approaches, namely BaseChatGPT-4, BaseChatGPT-3.5, BaseCodeLlama-13b, and BaseDeepSeek-33b, each directly utilizing basic LLMs to generate repairs using only the buggy function, associated error messages, and corresponding test cases with perfect FL information. We then apply \relrepair the same set of LLMs—specifically ChatGPT-4, ChatGPT-3.5, CodeLlama \cite{roziere2023code}, and DeepSeek \cite{deepseek2023coder} to assess the improvement gained by incorporating relevant information retrieval. The evaluation is conducted on the single-function repair benchmark provided by the Defects4J V1.2 dataset, enabling a systematic comparison of repair effectiveness in different models and configurations.

\subsubsection{\textbf{Results}}
Table~\ref{tab:RelRrepair vs. Basic LLMs} compares the effectiveness of \relrepair integrated with various LLMs (ChatGPT-4, ChatGPT-3.5, CodeLlama-13b, and DeepSeek-33b) against their basic LLM counterparts on the Defects4J V1.2 benchmark. The numbers highlighted in grey indicate the highest number of fixed bugs achieved for each project. The results indicate that the integration of \relrepair substantially improves the performance of bug fixing in all evaluated LLMs and projects. Notably, ChatGPT-4 combined with \relrepair achieves the highest total of 120 fixed bugs compared to 86 by its basic version, reflecting a significant improvement of approximately 37.2\%. Similarly, the integration with ChatGPT-3.5 (101 vs. 47), CodeLlama-13b (59 vs. 32), and DeepSeek-33b (68 vs. 34) demonstrates marked performance gains, with respective improvements of approximately 112.8\%, 87.5\%, and 105.9\%.

\begin{tcolorbox}[colback=white, colframe=black, arc=5pt, boxrule=0.8pt, left=2pt, right=2pt, top=2pt, bottom=2pt]
    \textbf{Findings to \textbf{RQ2}:} 
\textit{ Basic LLMs exhibit moderate performance in addressing single-function bugs within the Defects4J V1.2 dataset. In contrast, \relrepair consistently enhances the effectiveness of these LLMs, allowing them to successfully resolve a substantial number of bugs previously unaddressed by their basic counterparts.}
\end{tcolorbox}

\begin{table}[h]
    \centering
    \caption{Comparison of SigRepair and SnipRepair}
    \begin{tabular}{l|ccc}
        \toprule
        &  \multicolumn{3}{c}{\textbf{RelRepair}} \\
        \cmidrule(lr){2-4}
        & \textbf{BaseRepair} & \textbf{SigRepair} & \textbf{SnipRepair} \\
        \midrule
        Chart   & 5  & 1  & 4  \\
        Closure & 10 & 11 & 14  \\
        Lang    & 8  & 3  & 8   \\
        Math    & 18 & 0  & 10  \\
        Mockito & 4  & 0  & 1   \\
        Time    & 2  & 0  & 2   \\
        \midrule % Added line before RelRepair total
        
        \multirow{2}{*} {\textbf{Total Number}} & \textbf{47} & \textbf{15} & \textbf{39} \\  
        \cmidrule(lr){2-4}
        & \multicolumn{3}{c}{\textbf{101}} \\  
        \bottomrule
    \end{tabular}
    \label{tab:Comparison of SigRepair and SnipRepair}
\end{table}

\subsection{RQ3: Configurations of \relrepair}
\subsubsection{\textbf{Experimental Design}}
In this study, we systematically analyze various configurations of \relrepair to investigate the influence of its key components. RelRepair comprises two principal components: \whitecircled{1} SigRepair, a lightweight approach focused on retrieving relevant function signatures, and \whitecircled{2} SnipRepair, a more sophisticated strategy designed to retrieve relevant code snippets. Thus, in this research question, we comprehensively evaluate the distinct impacts of these two retrieval strategies on \relrepair's overall performance across two datasets: Defects4J V1.2 and ManySStuBs4J. Furthermore, we investigate three additional critical factors: (1) the importance of query rewriting in the retrieval process, (2) a comparative analysis of resource consumption between SigRepair and SnipRepair and (3) a discussion on the impact of parameter selection in SigRepair and SnipRepair. Given the significant cost incurred by repeatedly invoking the ChatGPT API during an extensive ablation study, we limit our evaluations for these latter two aspects to the 255 single-function bugs in Defects4J version 1.2. This targeted selection enables us to perform a focused and cost-effective assessment while ensuring meaningful and relevant insights into the effects of each evaluated parameter.

\subsubsection{\textbf{Results}}
\textbf{Analysis of Two Components}
According to the results presented in Table ~\ref{tab:Comparison of SigRepair and SnipRepair} and the provided Figure ~\ref{fig:manyststubs4j} analysis, several key observations emerge regarding the SigRepair and SnipRepair methods employed by \relrepair:

The evaluation of \relrepair demonstrates the distinct strengths and limitations of its two repair strategies. The SigRepair, designed as a lightweight and efficient approach, focuses on function modifications and additions without requiring deep code analysis, ensuring low computational cost. However, its effectiveness is limited when addressing complex bugs. Experimental results in Defects4J V1.2 show that SigRepair can only fix a narrow set of simpler bugs (e.g., 11 fixes in Closure and 3 fixes in Lang). In cases where SigRepair is insufficient, \relrepair escalates to SnipRepair, which retrieves semantically similar code snippets to enrich the contextual understanding of ChatGPT, allowing deeper code analysis and generating more accurate fixes. SnipRepair consistently exhibits broader applicability across various and complex bug types, significantly improving repair rates on challenging projects such as Closure (14 fixes), Math (10 fixes), and Lang (8 fixes).

Further analysis in ManySStuBs4J reinforces these findings. The SigRepair performs best in structured, function-level categories like \texttt{DIFFERENT\_METHOD\_SAME\_ARGS} (33.3\%) and \texttt{CHANGE\_IDENTIFIER} (13.3\%), where repairs mainly involve direct
modifications or replacements of function signatures. In contrast, SnipRepair surpasses SigRepair repair in categories requiring deeper understanding, such as \texttt{CHANGE\_MODIFIER} (30.0\%) and \texttt{ADD\_THROWS\_EXCEPTION} (30.0\%), where fixes mainly involve fix logic or illustrating function usage. These results highlight that while SigRepair offers fast and cost-effective fixes for simple cases, SnipRepair is crucial to address more intricate bugs that require more information.

\begin{table}[h]
    \centering
    \caption{Results of Query Rewriting.}
    \renewcommand{\arraystretch}{1.2}
    \begin{tabular}{lccccc}
        \hline
        \textbf{Rewriting} & \textbf{SigRepair} & \textbf{SnipRepair} & \textbf{\# Sum}\\ 
        \hline
        w & 15 & 39 & 54 \\ 
        w/o & 4 & 16 & 20\\ 
        \hline
    \end{tabular}
    \label{tab:ablation_study}
\end{table}
\textbf{Query Rewriting}
According to the results presented in Table ~\ref{tab:ablation_study}, several key observations can be made regarding the impact of query rewriting on \relrepair’s performance: (1) Query rewriting significantly influences the overall repair capability, markedly enhancing both SigRepair and SnipRepair. Specifically, when query rewriting is enabled (denoted as "w"), \relrepair successfully fixes a total of 54 bugs, whereas only 20 bugs are fixed without query rewriting ("w/o"), reflecting a considerable improvement. (2) The SigRepair particularly benefits from query rewriting, improving from 4 fixes without rewriting to 15 fixes with rewriting. Likewise, the SnipRepair approach shows substantial enhancement, increasing from 16 fixes without rewriting to 39 fixes with query rewriting enabled. \textit{(3)} These results highlight that refined queries provided by the rewriting process effectively clarify the repair context, significantly reducing ambiguity and enabling the model to generate more precise patches. Although query rewriting demonstrates overall effectiveness, its specific impact varies based on the nature of the bug. Complex defects requiring deeper analysis gain more pronounced benefits, as query refinement provides clearer instructions, while simpler function issues might see relatively limited improvement. 

\begin{figure}
    \centering
    \includegraphics[width=0.45\textwidth]{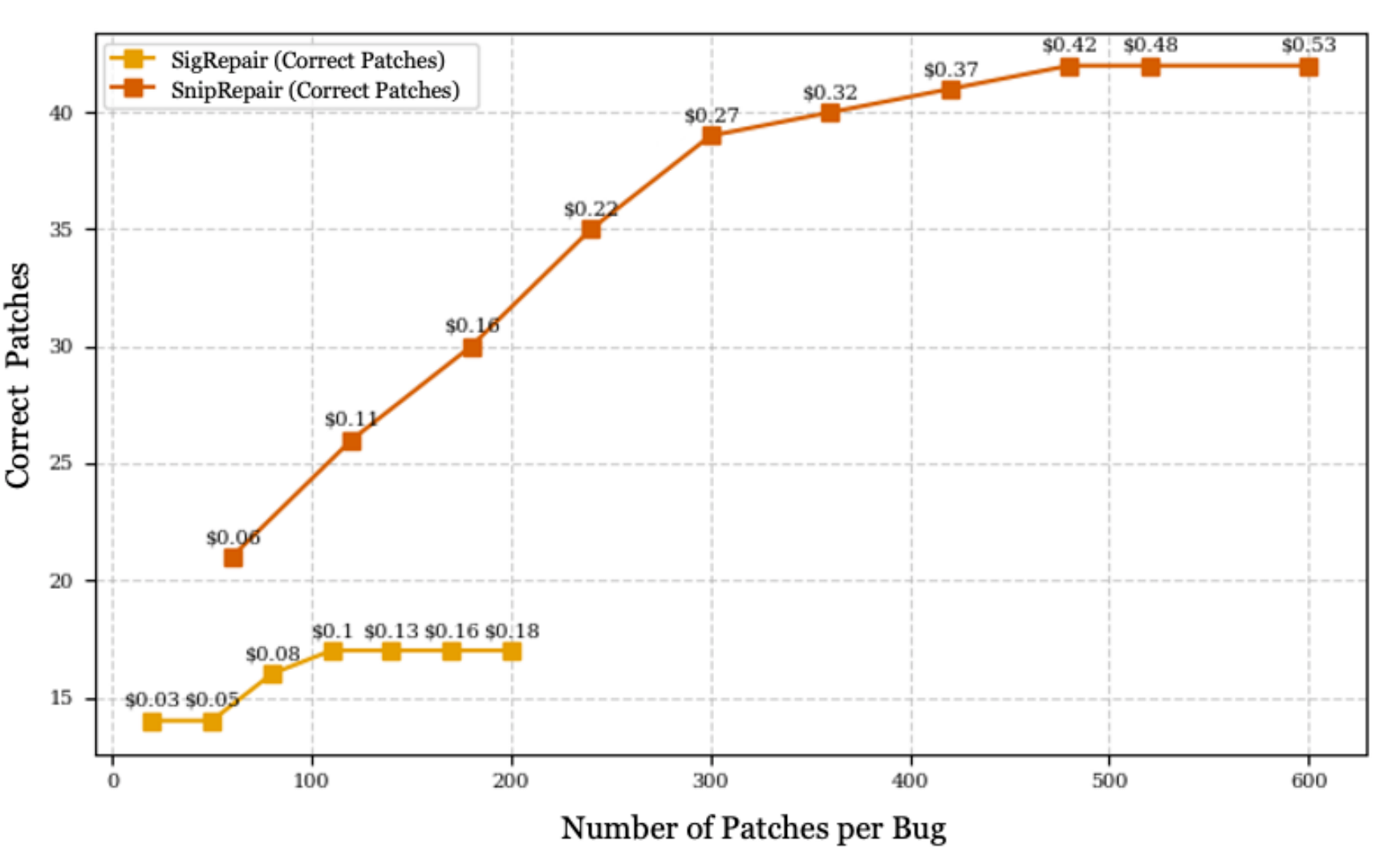}
    \caption{The cost of RelRepair with different numbers of patches}
    \label{fig:cost}
\end{figure}

\textbf{Resource Consumption}
According to the results presented in Fig.~\ref{fig:cost}, several key observations regarding the effectiveness and associated resource consumption of SigRepair and SnipRepair methods can be highlighted: (1) The SigRepair, designed as a lightweight solution targeting simpler bug patterns, achieves a moderate repair performance (up to 17 correct patches), rapidly reaching saturation despite increasing the number of generated patches. For instance, increasing patches from 80 to 200 yields minimal improvement (16→17), indicating performance limitations primarily due to design constraints rather than patch quantity. (2) In contrast, the SnipRepair, although incurring significantly higher computational costs (up to \$0.27 for 300 patches), exhibits consistent and substantial performance gains, increasing steadily from 21 correct fixes at 60 patches to 39 correct fixes at 300 patches. (3) These results underline that while SigRepair offer cost-effective but limited solutions for simpler bug types, SnipRepair, leveraging deeper contextual analysis, deliver notably greater accuracy and effectiveness at higher resource expenditures.

\begin{figure}
    \centering
    \includegraphics[width=0.45\textwidth]{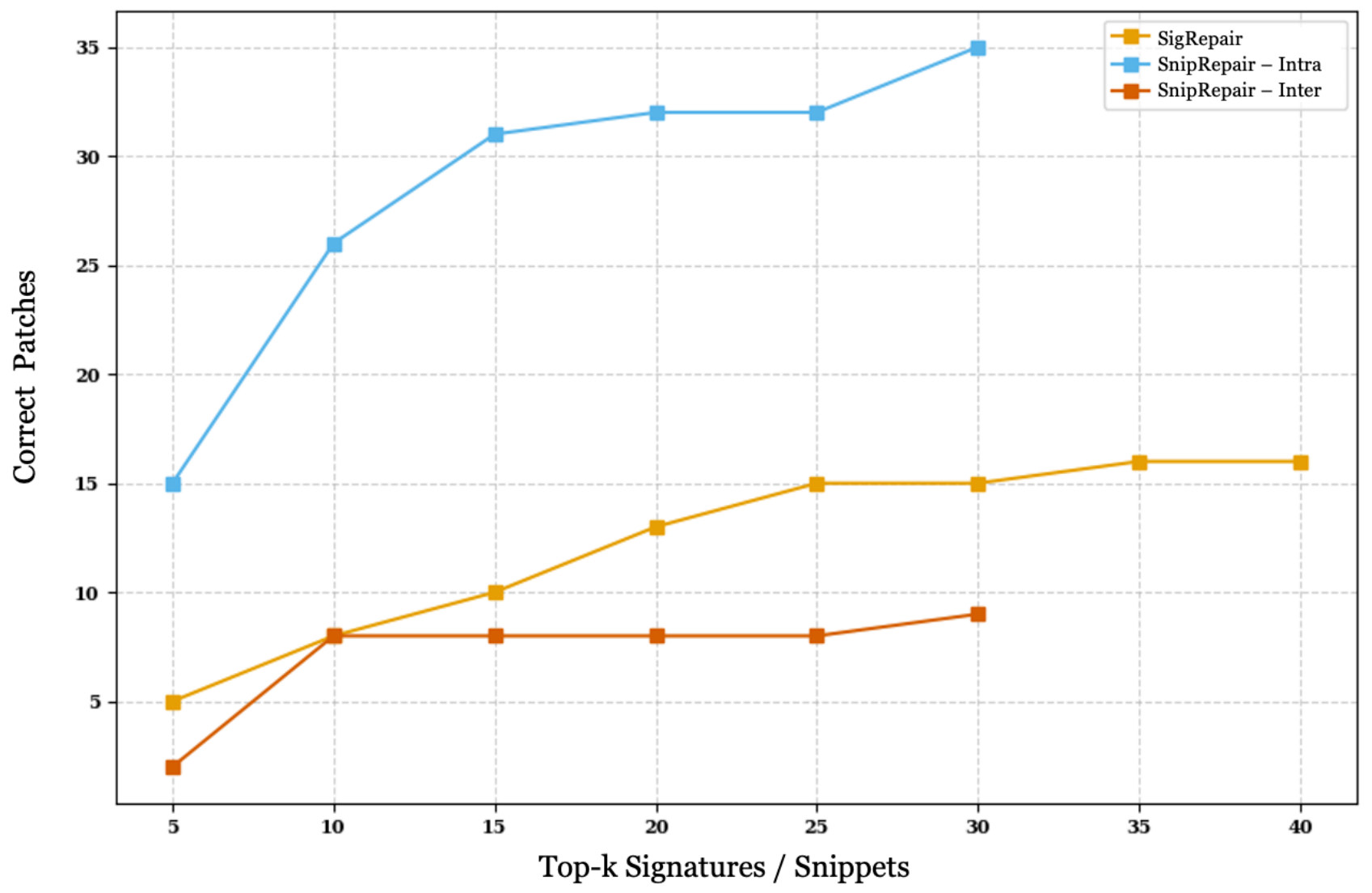}
    \caption{Performance of Parameter Selection}
    \label{fig:parameter}
\end{figure}

\textbf{Parameter Selection}
According to the results presented in Fig.~\ref{fig:parameter}, several important insights emerge regarding the influence of parameter selection for both SigRepair and SnipRepair. 
\begin{itemize}
    \item For SigRepair, the number of correct patches increases steadily as the number of retrieved function signatures grows, reaching 15 correct fixes at top-25. Beyond this point, further increases in retrieval depth yield only marginal improvements, with the correct patch count plateauing at 16 by top-35. This early saturation highlights SigRepair's lightweight nature: while it efficiently addresses a subset of bugs using only signature-level information, its reliance on function signatures inherently limits its repair capacity.
    \item In contrast, SnipRepair-Intra demonstrates a pronounced and sustained improvement, with correct patches increasing from 15 at top-5 to 31 at top-15. This substantial gain underscores the importance of leveraging intra-file code snippets, as local context often contains highly relevant patterns and implementation details crucial for accurate repair. Notably, this intra-file advantage aligns with our empirical observation that the majority of semantically relevant information for bug fixing is frequently found within the same file as the buggy function. As a result, prioritizing intra-file retrieval enables more precise and effective patch generation. However, increasing the number of retrieved intra-file code snippets beyond the top-15 yields diminishing returns, as the number of correct patches remains largely unchanged. Moreover, since each additional code snippet introduces an extra batch of candidate patches—specifically, 10 new patches per snippet—the computational and validation costs grow substantially.
    \item On the other hand, SnipRepair-Inter achieves only modest gains, with correct patches increasing from 2 at top-5 to 8 at top-15, and with little improvement beyond this point. While incorporating inter-file snippets offers some benefit—presumably by capturing broader project-level knowledge—its overall contribution remains relatively limited compared to intra-file retrieval. This finding further validates our retrieval strategy, which is designed to exhaust intra-file candidates before considering inter-file options.
\end{itemize}

\begin{tcolorbox}[colback=white, colframe=black, arc=5pt, boxrule=0.8pt, left=2pt, right=2pt, top=2pt, bottom=2pt]
    \textbf{Findings to \textbf{RQ3}:} 
\textit{  (1) Both SigRepair and SnipRepair significantly enhance \relrepair's bug-fixing capabilities, with SnipRepair notably more effective in handling complex bugs. (2) Query rewriting considerably improves \relrepair's performance, particularly benefiting SnipRepair by providing clearer context and instructions. (3) The SigRepair method is cost-efficient but limited in effectiveness, whereas the SnipRepair incurs higher costs but achieves substantially better repair accuracy. (4) The effectiveness of both SigRepair and SnipRepair depends on parameter selection. For both methods, increasing the number of retrieved signatures or code snippets beyond a certain threshold leads to higher computational costs while yielding diminishing returns in repair performance.}
\end{tcolorbox}

\section{THREATS TO VALIDITY}
\textbf{Internal}. The main internal threat arises from the manual validation performed to determine the correctness of plausible patches compared with reference developer fixes. To mitigate this concern, following established practices in previous research \cite{jiang2021cure, xia2023automated, xia2022less, ye2022selfapr, ye2022neural, zhu2021syntax}, we carefully examined each generated patch to ensure accurate validation. 

\textbf{External}. The main external threat
is the representativeness of the subjects used in our evaluation.
%relates to the generalizability of our evaluation results. 
To mitigate this concern, we performed evaluations on multiple datasets, including Defects4J V1.2 and ManySStuBs4J, covering a diverse range of real-world Java Maven projects. This broad evaluation framework supports the claim that RelRepair's improvements and effectiveness extend beyond isolated cases and demonstrate practical value across various software engineering scenarios.

\section{Related Work}
APR research has increasingly leveraged LLMs to effectively address diverse software defects. Recent notable advancements include ThinkRepair \cite{yin2024thinkrepair}, which enhances the analytical and reasoning capabilities of LLMs through Chain-of-Thought (CoT) methodologies. This allows it to grasp deeper semantic contexts of bugs and thus produce more accurate and contextually relevant patches. RAP-Gen \cite{wang2023rap} distinctly leverages historical bug-fix pairs, employing a hybrid retrieval mechanism that combines sparse and dense retrievers. This dual approach effectively supports both lexical and semantic matching, enabling RAP-Gen to operate efficiently across various programming languages directly from raw source code. AutoCodeRover \cite{zhang2404autocoderover} uniquely integrates the structural aspects of software through Abstract Syntax Trees (ASTs), improving understanding of the root causes of the defect through iterative and structural code analyses. This results in highly precise and context-sensitive repairs. Conversational approaches, exemplified by ChatRepair \cite{xia2023keep}, utilize interactive dialogues between models and users, creating feedback loops that iteratively refine repair patches. This conversational dynamic enables continuous user-driven improvements, bridging automated repair with manual debugging methods. Muplor takes a complementary approach by providing multi-granularity patch generation capabilities. By addressing bug repairs at multiple levels of granularity, Muplor \cite{lin2024one} efficiently handles the complex and variable nature of real-world software defects.

In contrast, our proposed approach, \textit{RelRepair}, significantly improves the quality and accuracy of generated patches by explicitly retrieving the relevant function signatures and code snippets. This explicit semantic retrieval systematically improves both the accuracy and effectiveness of patches generated by LLMs. Unlike RAP-Gen, which relies on historical bug-fix patterns, or AutoCodeRover’s AST-based structural analyses, \textit{RelRepair} utilizes retrieved information independently of historical patterns or structural analysis. Moreover, compared to ThinkRepair’s analytical reasoning approach via CoT, \textit{RelRepair} prioritizes explicit semantic retrieval strategies. Together, these varied methods reflect the strategic efforts within LLM-based APRs to enhance the quality and accuracy of generated patches, emphasizing the importance of retrieval, structural analysis, conversational interaction, multi-granularity generation, and template-driven fine-tuning in addressing complex software repair challenges.

\section{CONCLUSION}
We introduced RelRepair, a novel approach that retrieves relevant code to enhance automated program repair, leveraging a specialized RAG strategy for program repair tasks. Specifically, \relrepair employs RAG to identify and incorporate relevant function signatures and code snippets, to guide LLMs to generate more accurate and informed patches. Evaluation of two widely used benchmarks, Defects4J V1.2 and ManySStuBs4J, demonstrates the effectiveness of \relrepair in program repair. Specifically, RelRepair successfully repairs 101 bugs in Defects4J V1.2, including 28 unique repairs not achieved by other tools, and substantially improves the overall repair rate in ManySStuBs4J from 31.2\% to 48.3\%. These findings confirm that retrieving relevant code from the project codebase is essential for enhancing APR and highlight the practical utility of \relrepair in real-world bug-fixing problems.

\bibliographystyle{IEEEtran}
\balance
\bibliography{sample-base}

% Generated by IEEEtran.bst, version: 1.14 (2015/08/26)
\begin{thebibliography}{10}
\providecommand{\url}[1]{#1}
\csname url@samestyle\endcsname
\providecommand{\newblock}{\relax}
\providecommand{\bibinfo}[2]{#2}
\providecommand{\BIBentrySTDinterwordspacing}{\spaceskip=0pt\relax}
\providecommand{\BIBentryALTinterwordstretchfactor}{4}
\providecommand{\BIBentryALTinterwordspacing}{\spaceskip=\fontdimen2\font plus
\BIBentryALTinterwordstretchfactor\fontdimen3\font minus \fontdimen4\font\relax}
\providecommand{\BIBforeignlanguage}[2]{{%
\expandafter\ifx\csname l@#1\endcsname\relax
\typeout{** WARNING: IEEEtran.bst: No hyphenation pattern has been}%
\typeout{** loaded for the language `#1'. Using the pattern for}%
\typeout{** the default language instead.}%
\else
\language=\csname l@#1\endcsname
\fi
#2}}
\providecommand{\BIBdecl}{\relax}
\BIBdecl

\bibitem{long2016analysis}
F.~Long and M.~Rinard, ``An analysis of the search spaces for generate and validate patch generation systems,'' in \emph{Proceedings of the 38th International Conference on Software Engineering}, 2016, pp. 702--713.

\bibitem{le2011genprog}
C.~Le~Goues, T.~Nguyen, S.~Forrest, and W.~Weimer, ``Genprog: A generic method for automatic software repair,'' \emph{Ieee transactions on software engineering}, vol.~38, no.~1, pp. 54--72, 2011.

\bibitem{le2016history}
X.~B.~D. Le, D.~Lo, and C.~Le~Goues, ``History driven program repair,'' in \emph{2016 IEEE 23rd international conference on software analysis, evolution, and reengineering (SANER)}, vol.~1.\hskip 1em plus 0.5em minus 0.4em\relax IEEE, 2016, pp. 213--224.

\bibitem{wen2018context}
M.~Wen, J.~Chen, R.~Wu, D.~Hao, and S.-C. Cheung, ``Context-aware patch generation for better automated program repair,'' in \emph{Proceedings of the 40th international conference on software engineering}, 2018, pp. 1--11.

\bibitem{demarco2014automatic}
F.~DeMarco, J.~Xuan, D.~Le~Berre, and M.~Monperrus, ``Automatic repair of buggy if conditions and missing preconditions with smt,'' in \emph{Proceedings of the 6th international workshop on constraints in software testing, verification, and analysis}, 2014, pp. 30--39.

\bibitem{long2015staged}
F.~Long and M.~Rinard, ``Staged program repair with condition synthesis,'' in \emph{Proceedings of the 2015 10th Joint Meeting on Foundations of Software Engineering}, 2015, pp. 166--178.

\bibitem{mechtaev2016angelix}
S.~Mechtaev, J.~Yi, and A.~Roychoudhury, ``Angelix: Scalable multiline program patch synthesis via symbolic analysis,'' in \emph{Proceedings of the 38th international conference on software engineering}, 2016, pp. 691--701.

\bibitem{le2017s3}
X.-B.~D. Le, D.-H. Chu, D.~Lo, C.~Le~Goues, and W.~Visser, ``S3: syntax-and semantic-guided repair synthesis via programming by examples,'' in \emph{Proceedings of the 2017 11th Joint Meeting on Foundations of Software Engineering}, 2017, pp. 593--604.

\bibitem{martinez2016astor}
M.~Martinez and M.~Monperrus, ``Astor: A program repair library for java,'' in \emph{Proceedings of the 25th international symposium on software testing and analysis}, 2016, pp. 441--444.

\bibitem{hua2018sketchfix}
J.~Hua, M.~Zhang, K.~Wang, and S.~Khurshid, ``Sketchfix: a tool for automated program repair approach using lazy candidate generation,'' in \emph{Proceedings of the 2018 26th ACM Joint Meeting on European Software Engineering Conference and Symposium on the Foundations of Software Engineering}, 2018, pp. 888--891.

\bibitem{ghanbari2019practical}
A.~Ghanbari, S.~Benton, and L.~Zhang, ``Practical program repair via bytecode mutation,'' in \emph{Proceedings of the 28th ACM SIGSOFT International Symposium on Software Testing and Analysis}, 2019, pp. 19--30.

\bibitem{liu2019tbar}
K.~Liu, A.~Koyuncu, D.~Kim, and T.~F. Bissyand{\'e}, ``Tbar: Revisiting template-based automated program repair,'' in \emph{Proceedings of the 28th ACM SIGSOFT international symposium on software testing and analysis}, 2019, pp. 31--42.

\bibitem{liu2019avatar}
------, ``Avatar: Fixing semantic bugs with fix patterns of static analysis violations,'' in \emph{2019 IEEE 26th International Conference on Software Analysis, Evolution and Reengineering (SANER)}.\hskip 1em plus 0.5em minus 0.4em\relax IEEE, 2019, pp. 1--12.

\bibitem{chen2019sequencer}
Z.~Chen, S.~Kommrusch, M.~Tufano, L.-N. Pouchet, D.~Poshyvanyk, and M.~Monperrus, ``Sequencer: Sequence-to-sequence learning for end-to-end program repair,'' \emph{IEEE Transactions on Software Engineering}, vol.~47, no.~9, pp. 1943--1959, 2019.

\bibitem{jiang2021cure}
N.~Jiang, T.~Lutellier, and L.~Tan, ``Cure: Code-aware neural machine translation for automatic program repair,'' in \emph{2021 IEEE/ACM 43rd International Conference on Software Engineering (ICSE)}.\hskip 1em plus 0.5em minus 0.4em\relax IEEE, 2021, pp. 1161--1173.

\bibitem{li2020dlfix}
Y.~Li, S.~Wang, and T.~N. Nguyen, ``Dlfix: Context-based code transformation learning for automated program repair,'' in \emph{Proceedings of the ACM/IEEE 42nd international conference on software engineering}, 2020, pp. 602--614.

\bibitem{lutellier2020coconut}
T.~Lutellier, H.~V. Pham, L.~Pang, Y.~Li, M.~Wei, and L.~Tan, ``Coconut: combining context-aware neural translation models using ensemble for program repair,'' in \emph{Proceedings of the 29th ACM SIGSOFT international symposium on software testing and analysis}, 2020, pp. 101--114.

\bibitem{kolak2022patch}
S.~D. Kolak, R.~Martins, C.~Le~Goues, and V.~J. Hellendoorn, ``Patch generation with language models: Feasibility and scaling behavior,'' in \emph{Deep Learning for Code Workshop}, 2022.

\bibitem{prenner2022can}
J.~A. Prenner, H.~Babii, and R.~Robbes, ``Can openai's codex fix bugs? an evaluation on quixbugs,'' in \emph{Proceedings of the Third International Workshop on Automated Program Repair}, 2022, pp. 69--75.

\bibitem{xia2023automated}
C.~S. Xia, Y.~Wei, and L.~Zhang, ``Automated program repair in the era of large pre-trained language models,'' in \emph{2023 IEEE/ACM 45th International Conference on Software Engineering (ICSE)}.\hskip 1em plus 0.5em minus 0.4em\relax IEEE, 2023, pp. 1482--1494.

\bibitem{schulman2022chatgpt}
J.~Schulman, B.~Zoph, J.~Hilton, C.~Kim, J.~Menick, J.~Weng, J.~F.~C. Uribe, L.~Fedus, L.~Metz, M.~Pokorny, R.~G. Lopes, S.~Zhao, A.~Vijayvergiya, E.~Sigler, A.~Perelman, C.~Voss, M.~Heaton, J.~Parish, D.~Cummings, R.~Nayak, V.~Balcom, D.~Schnurr, T.~Kaftan, C.~Hallacy, N.~Turley, N.~Deutsch, V.~Goel, J.~Ward, A.~Konstantinidis, W.~Zaremba, L.~Ouyang, L.~Bogdonoff, J.~Gross, D.~Medina, S.~Yoo, T.~Lee, R.~Lowe, D.~Mossing, J.~Huizinga, R.~Jiang, C.~Wainwright, D.~Almeida, S.~Lin, M.~Zhang, K.~Xiao, K.~Slama, S.~Bills, A.~Gray, J.~Leike, J.~Pachocki, P.~Tillet, S.~Jain, G.~Brockman, and N.~Ryder, ``Chatgpt: Optimizing language models for dialogue,'' \url{https://openai.com/blog/chatgpt/}, 2022.

\bibitem{xia2023keep}
C.~S. Xia and L.~Zhang, ``Keep the conversation going: Fixing 162 out of 337 bugs for \$0.42 each using chatgpt,'' \emph{arXiv preprint arXiv:2304.00385}, 2023.

\bibitem{yin2024thinkrepair}
X.~Yin, C.~Ni, S.~Wang, Z.~Li, L.~Zeng, and X.~Yang, ``Thinkrepair: Self-directed automated program repair,'' in \emph{Proceedings of the 33rd ACM SIGSOFT International Symposium on Software Testing and Analysis}, 2024, pp. 1274--1286.

\bibitem{lewis2020retrieval}
P.~Lewis, E.~Perez, A.~Piktus, F.~Petroni, V.~Karpukhin, N.~Goyal, H.~K{\"u}ttler, M.~Lewis, W.-t. Yih, T.~Rockt{\"a}schel \emph{et~al.}, ``Retrieval-augmented generation for knowledge-intensive nlp tasks,'' \emph{Advances in neural information processing systems}, vol.~33, pp. 9459--9474, 2020.

\bibitem{just2014defects4j}
R.~Just, D.~Jalali, and M.~D. Ernst, ``Defects4j: A database of existing faults to enable controlled testing studies for java programs,'' in \emph{Proceedings of the 2014 international symposium on software testing and analysis}, 2014, pp. 437--440.

\bibitem{karampatsis2020often}
R.-M. Karampatsis and C.~Sutton, ``How often do single-statement bugs occur? the manysstubs4j dataset,'' in \emph{Proceedings of the 17th International Conference on Mining Software Repositories}, 2020, pp. 573--577.

\bibitem{brown2020language}
T.~Brown, B.~Mann, N.~Ryder, M.~Subbiah, J.~D. Kaplan, P.~Dhariwal, A.~Neelakantan, P.~Shyam, G.~Sastry, A.~Askell \emph{et~al.}, ``Language models are few-shot learners,'' \emph{Advances in neural information processing systems}, vol.~33, pp. 1877--1901, 2020.

\bibitem{radford2018improving}
A.~Radford, K.~Narasimhan, T.~Salimans, I.~Sutskever \emph{et~al.}, ``Improving language understanding by generative pre-training,'' 2018.

\bibitem{liu2023pre}
P.~Liu, W.~Yuan, J.~Fu, Z.~Jiang, H.~Hayashi, and G.~Neubig, ``Pre-train, prompt, and predict: A systematic survey of prompting methods in natural language processing,'' \emph{ACM computing surveys}, vol.~55, no.~9, pp. 1--35, 2023.

\bibitem{ziegler2019fine}
D.~M. Ziegler, N.~Stiennon, J.~Wu, T.~B. Brown, A.~Radford, D.~Amodei, P.~Christiano, and G.~Irving, ``Fine-tuning language models from human preferences,'' \emph{arXiv preprint arXiv:1909.08593}, 2019.

\bibitem{christiano2017deep}
P.~F. Christiano, J.~Leike, T.~Brown, M.~Martic, S.~Legg, and D.~Amodei, ``Deep reinforcement learning from human preferences,'' \emph{Advances in neural information processing systems}, vol.~30, 2017.

\bibitem{ouyang2022training}
L.~Ouyang, J.~Wu, X.~Jiang, D.~Almeida, C.~Wainwright, P.~Mishkin, C.~Zhang, S.~Agarwal, K.~Slama, A.~Ray \emph{et~al.}, ``Training language models to follow instructions with human feedback,'' \emph{Advances in neural information processing systems}, vol.~35, pp. 27\,730--27\,744, 2022.

\bibitem{schulman2017proximal}
J.~Schulman, F.~Wolski, P.~Dhariwal, A.~Radford, and O.~Klimov, ``Proximal policy optimization algorithms,'' \emph{arXiv preprint arXiv:1707.06347}, 2017.

\bibitem{gehring2017convolutional}
J.~Gehring, M.~Auli, D.~Grangier, D.~Yarats, and Y.~N. Dauphin, ``Convolutional sequence to sequence learning,'' in \emph{International conference on machine learning}.\hskip 1em plus 0.5em minus 0.4em\relax PMLR, 2017, pp. 1243--1252.

\bibitem{kandpal2023large}
N.~Kandpal, H.~Deng, A.~Roberts, E.~Wallace, and C.~Raffel, ``Large language models struggle to learn long-tail knowledge,'' in \emph{International Conference on Machine Learning}.\hskip 1em plus 0.5em minus 0.4em\relax PMLR, 2023, pp. 15\,696--15\,707.

\bibitem{li2022survey}
H.~Li, Y.~Su, D.~Cai, Y.~Wang, and L.~Liu, ``A survey on retrieval-augmented text generation,'' \emph{arXiv preprint arXiv:2202.01110}, 2022.

\bibitem{gao2023retrieval}
Y.~Gao, Y.~Xiong, X.~Gao, K.~Jia, J.~Pan, Y.~Bi, Y.~Dai, J.~Sun, H.~Wang, and H.~Wang, ``Retrieval-augmented generation for large language models: A survey,'' \emph{arXiv preprint arXiv:2312.10997}, vol.~2, 2023.

\bibitem{lin2017quixbugs}
D.~Lin, J.~Koppel, A.~Chen, and A.~Solar-Lezama, ``Quixbugs: A multi-lingual program repair benchmark set based on the quixey challenge,'' in \emph{Proceedings Companion of the 2017 ACM SIGPLAN international conference on systems, programming, languages, and applications: software for humanity}, 2017, pp. 55--56.

\bibitem{xia2022less}
C.~S. Xia and L.~Zhang, ``Less training, more repairing please: revisiting automated program repair via zero-shot learning,'' in \emph{Proceedings of the 30th ACM Joint European Software Engineering Conference and Symposium on the Foundations of Software Engineering}, 2022, pp. 959--971.

\bibitem{lin2024one}
B.~Lin, S.~Wang, M.~Wen, L.~Chen, and X.~Mao, ``One size does not fit all: Multi-granularity patch generation for better automated program repair,'' in \emph{Proceedings of the 33rd ACM SIGSOFT International Symposium on Software Testing and Analysis}, 2024, pp. 1554--1566.

\bibitem{wang2023rap}
W.~Wang, Y.~Wang, S.~Joty, and S.~C. Hoi, ``Rap-gen: Retrieval-augmented patch generation with codet5 for automatic program repair,'' in \emph{Proceedings of the 31st ACM Joint European Software Engineering Conference and Symposium on the Foundations of Software Engineering}, 2023, pp. 146--158.

\bibitem{gilardi2023chatgpt}
F.~Gilardi, M.~Alizadeh, and M.~Kubli, ``Chatgpt outperforms crowd workers for text-annotation tasks,'' \emph{Proceedings of the National Academy of Sciences}, vol. 120, no.~30, p. e2305016120, 2023.

\bibitem{zhu2021syntax}
Q.~Zhu, Z.~Sun, Y.-a. Xiao, W.~Zhang, K.~Yuan, Y.~Xiong, and L.~Zhang, ``A syntax-guided edit decoder for neural program repair,'' in \emph{Proceedings of the 29th ACM joint meeting on European software engineering conference and symposium on the foundations of software engineering}, 2021, pp. 341--353.

\bibitem{ye2022neural}
H.~Ye, M.~Martinez, and M.~Monperrus, ``Neural program repair with execution-based backpropagation,'' in \emph{Proceedings of the 44th international conference on software engineering}, 2022, pp. 1506--1518.

\bibitem{roziere2023code}
B.~Roziere, J.~Gehring, F.~Gloeckle, S.~Sootla, I.~Gat, X.~E. Tan, Y.~Adi, J.~Liu, R.~Sauvestre, T.~Remez \emph{et~al.}, ``Code llama: Open foundation models for code,'' \emph{arXiv preprint arXiv:2308.12950}, 2023.

\bibitem{deepseek2023coder}
\BIBentryALTinterwordspacing
D.~AI, ``Deepseek coder: Let the code write itself,'' 2023, accessed: 2025-04-15. [Online]. Available: \url{https://github.com/deepseek-ai/DeepSeek-Coder}
\BIBentrySTDinterwordspacing

\bibitem{ye2022selfapr}
H.~Ye, M.~Martinez, X.~Luo, T.~Zhang, and M.~Monperrus, ``Selfapr: Self-supervised program repair with test execution diagnostics,'' in \emph{Proceedings of the 37th IEEE/ACM International Conference on Automated Software Engineering}, 2022, pp. 1--13.

\bibitem{zhang2404autocoderover}
Y.~Zhang, H.~Ruan, Z.~Fan, and A.~Roychoudhury, ``Autocoderover: Autonomous program improvement, 2024,'' \emph{URL https://arxiv. org/abs/2404.05427}.

\end{thebibliography}
\end{document}